\documentclass[aps,pra,reprint,amsmath,amssymb,groupedaddress,showpacs]{revtex4-1} 
\usepackage{graphicx}
\usepackage{float}
\usepackage{dcolumn} 
\usepackage{bm}      
\usepackage[bookmarks=false]{hyperref}
\usepackage[font={small}]{caption}    
\usepackage{enumitem}
\usepackage{textcomp}
\hypersetup{
    unicode=false,          
    pdftoolbar=true,        
    pdfmenubar=true,        
    pdffitwindow=true,      
    pdfstartview={},        
    pdftitle={Explicit Inverse Confluent Vandermonde Matrices with Applications to Exponential Quantum Operators}, 
    pdfauthor={Samuel R. Hedemann},
    pdfsubject={Operator Functions in Quantum Mechanics},
    pdfcreator={},          
    pdfproducer={},         
    pdfkeywords={Vandermonde,} {Confluent Vandermonde Inverse,} {Matrix Inverse,} {Exponential Operator,} {Operator Function,} {Cayley-Hamilton Theorem,} {Elementary Symmetric Polynomials}, 
    pdfnewwindow=true,      
    colorlinks=true,        
    linkcolor=black,        
    citecolor=black,        
    filecolor=black,        
    urlcolor=black          
}
\newcommand{\Sec}[1]{\hyperref[sec:#1]{Sec.{\kern 2pt}\ref*{sec:#1}}}
\newcommand{\Section}[1]{\hyperref[sec:#1]{Section~\ref*{sec:#1}}}
\newcommand{\Fig}[2][]{\hyperref[fig:#2]{Fig.{\kern 2pt}\ref*{fig:#2}#1}}
\newcommand{\Figure}[2][]{\hyperref[fig:#2]{Figure~\ref*{fig:#2}#1}}
\newcommand{\App}[1]{\hyperref[sec:App.#1]{App.{\kern 2pt}\ref*{sec:App.#1}}}
\newcommand{\Appendix}[1]{\hyperref[sec:App.#1]{Appendix~\ref*{sec:App.#1}}}
\newcommand{\Eq}[1]{\hyperref[eq:#1]{(\ref*{eq:#1})}}
\newcommand{\Eqs}[2]{\hyperref[eq:#1]{(\ref*{eq:#1}--\ref*{eq:#2})}}
\newcommand{\Table}[2][]{\hyperref[tab:#2]{Table~\ref*{tab:#2}#1}}
\newcommand{\shiftmath}[2]{\textnormal{\raisebox{#1}[#1][#1]{$#2$}}}
\newcommand{\hsp}[1]{{\kern #1pt}}
\begin{document}
\title{Explicit Inverse Confluent Vandermonde Matrices with Applications to Exponential Quantum Operators}
\author{Samuel R.{\kern 3.4pt}Hedemann}
\noaffiliation
\date{\today}
\begin{abstract}
The Cayley-Hamilton problem of expressing functions of matrices in terms of only their eigenvalues is well-known to simplify to finding the inverse of the confluent Vandermonde matrix.  Here, we give a highly compact formula for the inverse of any matrix, and apply it to the confluent Vandermonde matrix, achieving in a single equation what has only been achieved by long iterative algorithms until now.  As a prime application, we use this result to get a simple formula for explicit exponential operators in terms of only their eigenvalues, with an emphasis on application to finite discrete quantum systems with time dependence.  This powerful result permits explicit solutions to all Schr{\"o}dinger and von Neumann equations for time-commuting Hamiltonians, and explicit solutions to any degree of approximation in the non-time-commuting case.  The same methods can be extended to general finite discrete open systems to get explicit quantum operations for time evolution using effective joint systems, and the exact solution of all finite discrete Baker-Campbell-Hausdorff formulas.
\end{abstract}
\maketitle
\section{\label{sec:I}Introduction and Main Results}
An important problem encountered in many situations is to find an exact expression for any analytic function $f(x)\equiv\sum\nolimits_{j=0}^{\infty}a_{j}x^{j}$ of an $n\times n$ matrix $A$ as $f(A)$ in terms of only the eigenvalues of $A$. While this problem has already been solved through the Cayley-Hamilton theorem \cite[]{Ham1,Ham2,Ham3,Ham4,Cay1,Frob,Cay2}, its solution requires an explicit form of the inverse confluent Vandermonde matrix.  In this work, we present a compact, exact form for the inverse confluent Vandermonde, thus explicitly solving all analytical functions of $n\times n$ operators as $f(A)$. While many others have found solutions to this problem \cite[]{Gaut,HoPa,LuRo,HoHo,SoMo}, the solution presented here is more compact and easily implemented.

As our focus for applications, we compute exponential operators in \textit{all} cases of finite-dimensional $A$, including time-dependent and non-time-commuting $A$.  Exponential operators are of prime importance in many fields, the simplest cases arising from the equation of motion $\partial _t \mathbf{x}(t){\kern -0.8pt} ={\kern -0.8pt} A\mathbf{x}(t)$,{\kern -0.8pt} where{\kern -0.8pt} $\partial _t  {\kern -0.8pt}\equiv{\kern -0.8pt} \frac{\partial }{{\partial t}}$,{\kern -0.8pt} with{\kern -0.8pt} solution{\kern -0.8pt} $\mathbf{x}(t) {\kern -0.8pt}\equiv {\kern -0.8pt}\mathbf{x}(t,t_0 )${\kern -0.8pt} subject{\kern -0.8pt} to{\kern -0.8pt} initial{\kern -0.8pt} condition{\kern -0.8pt} $\mathbf{x}(t_0 )$,{\kern -0.8pt} and{\kern -0.8pt} operator{\kern -0.8pt} $A$,{\kern -0.8pt} represented{\kern -0.8pt} by{\kern -0.8pt} a{\kern -0.8pt} complex-valued{\kern -0.8pt} $n{\kern -0.8pt}\times{\kern -0.8pt} n${\kern -0.8pt} matrix.{\kern -0.8pt} For{\kern -0.8pt} constant{\kern -0.8pt} $A$,{\kern -0.8pt} the{\kern -0.9pt} well-known{\kern -0.9pt} solution{\kern -0.9pt} is{\kern -0.9pt} $\mathbf{x}(t){\kern -0.9pt} ={\kern -0.9pt} e^{\shiftmath{-0.5pt}{(t - t_0 )A}} \mathbf{x}(t_0 )$,{\kern -0.9pt} thus{\kern -0.9pt} requiring{\kern -0.9pt} exponential{\kern -0.9pt} operator{\kern -0.9pt} \smash{$e^{\shiftmath{-0.5pt}{(t - t_0 )A}}$}\!.{\kern -1.0pt} A{\kern -0.9pt} key{\kern -0.9pt} example{\kern -0.9pt} is{\kern -0.9pt} the{\kern -0.8pt} Schr{\"o}dinger{\kern -0.8pt} equation{\kern -0.8pt} {$\partial _t |\psi (t)\rangle  {\kern -0.8pt}= {\kern -2pt} -{\kern -0.4pt} \frac{i}{\hbar }H|\psi (t)\rangle$} \cite[]{Sch1,Sch2}.{\kern 8pt}

For the most general operator-function problem, given an $n\times n$ square matrix $A$, with $M$ distinct eigenvalues \smash{$\bm{\lambda}\equiv(\lambda_{1},\ldots,\lambda_{M})$} where $1\leq M\leq n$, so that \smash{$M\equiv\text{dim}(\bm{\lambda})$}, with multiplicities \smash{$\bm{\mu}\equiv(\mu_{1},\ldots,\mu_{M})$} such that $\sum\nolimits_{\shiftmath{0.3pt}{k\!=\!1}}^{\shiftmath{-0.7pt}{M}}\mu_{k}=n$, for any analytic function $f(x)\equiv\sum\nolimits_{j=0}^{\infty}a_{j}x^{\shiftmath{-0.3pt}{j}}$, we can obtain the operator function $f(A)$ as
\begin{equation}
f(A) = \sum\limits_{k = 1}^n {b_k A^{k - 1} },
\label{eq:1}
\end{equation}
with $A^0 =I$, and scalars $b_k$ given by elements of vector
\begin{equation}
\mathbf{b} = V^{ - 1} \widetilde{\mathbf{f}},\;\;\;\text{where}\;\;\;\widetilde{f}_{c_{\alpha,\beta}}\equiv f^{(\beta -1)}(x)|_{\lambda_{\alpha}},
\label{eq:2}
\end{equation}
where $\mathbf{b}\equiv(b_{\shiftmath{0.7pt}{1}},\ldots,b_{\shiftmath{0.7pt}{n}})^{T}$,  $\widetilde{\mathbf{f}}\equiv(\widetilde{f}_{1},\ldots,\widetilde{f}_{n})^{T}$, and \smash{$f^{(m)} (x)|_{\shiftmath{0.7pt}{y}} {\kern -0.6pt} \equiv {\kern -0.6pt}({\textstyle{{d^{\shiftmath{-2pt}{m}} {\kern 0.4pt}f{\kern 0.6pt}(x)} \over {dx^m }}})|_{\shiftmath{0.7pt}{x{\kern -1pt} ={\kern -1pt} y}}$}, with \textit{indical register function} \smash{$c_{\alpha ,\beta }  \equiv (\sum\nolimits_{j = 1}^{\shiftmath{-0.7pt}{\alpha{\kern -1pt} -{\kern -1pt} 1}} {\mu _j }){\kern -0.5pt}+{\kern -0.5pt}\beta$}{\kern -0.5pt} for{\kern -0.5pt} $\alpha  {\kern -0.5pt}\in{\kern -0.5pt} 1, \ldots ,M${\kern -0.5pt} and{\kern -0.5pt} $\beta  {\kern -0.5pt}\in{\kern -0.5pt} 1, \ldots ,\mu _\alpha$,{\kern -1pt} and{\kern -1pt} \textit{confluent{\kern -1pt} Vandermonde{\kern -1pt} matrix}{\kern -1pt} $V$,{\kern -1pt} with{\kern -1pt} elements\rule{0pt}{9.6pt}
\begin{equation}
V_{c_{\alpha,\beta} ,d}\equiv\partial_{\lambda_{\alpha}}^{\beta-1}\lambda_{\alpha}^{d-1}=\left\{ {\begin{array}{*{20}l}
   0; & {d < \beta }  \\
   {\frac{{(d - 1)!}}{{(d - \beta )!}}\lambda _{\alpha} ^{d - \beta }; } & {d \ge \beta .}  \\
\end{array}} \right.
\label{eq:3}
\end{equation}

Due to the appearance of $V^{-1}$ in \Eq{2}, the problem of \Eq{1} reduces to finding an explicit form for $V^{-1}$.  Several formulas for this have already been discovered \cite[]{Gaut,HoPa,LuRo,HoHo,SoMo}, but they are generally recursive, algorithmic, and not compact, often taking many pages to describe. Here, we give a relatively simple expression for $V^{-1}$ in terms of $V$ as
\begin{equation}
V^{ - 1}{\kern -0.5pt}  = \frac{{\frac{{( - 1)^{n + 1} }}{{n + 1}}{\kern -0.5pt}\sum\limits_{k = 1}^n {V^{k - 1} {\kern -0.5pt}\sum\limits_{j = 0}^n {e^{i\frac{{2\pi }}{{n + 1}}jk} \det (e^{ - i\frac{{2\pi }}{{n + 1}}j} I {\kern -0.5pt}-{\kern -0.5pt} V)} } }}{{\det (V)}},
\label{eq:4}
\end{equation}
where $i{\kern -0.5pt} \equiv{\kern -0.5pt} \sqrt{-1}$, and the dependency on eigenvalues of $A$ is seen by putting \Eq{3} into \Eq{4}.  See \App{A} for a derivation of \Eq{4}. Note that \Eq{4} holds for \textit{any} $n$-level square matrix $V$ for which $\text{det}(V){\kern -0.5pt} \neq {\kern -0.5pt} 0$, so it is also a general closed form for the inverse of a matrix.  For the confluent Vandermonde $V$, $\text{det}(V){\kern -0.5pt} \neq {\kern -0.5pt} 0$ always, and we can get $V^{-1}$ in terms of either \textit{elements} of $V$ or eigenvalues $\bm{\lambda}$ of $A$, as shown in \App{B}. See \App{C} for a symbolic-software-friendly $V^{-1}$ without the complex exponentials.

The \textit{benefits} of \Eq{4} over other matrix inverse forms are that \Eq{4} does \textit{not} require recursive determinants of submatrices [see \Eq{7}], and it has no constrained indices, unlike other forms such as those involving Bell polynomials.  Thus, \Eq{4} may be easier to use when we need to see the eigenvalues $\bm{\lambda}$ of $A$ or the elements of $V$ in $V^{-1}$ explicitly.

This closed form in \Eq{4} is just the definition of a matrix inverse with discrete Fourier orthogonality applied to the Cayley-Hamilton theorem's formula \cite[]{Ham1,Ham2,Ham3,Ham4,Cay1,Frob,Cay2} (see \App{A}) for the adjugate of a matrix. Compartmentalizing,
\begin{equation}
V^{ - 1}  = \frac{\text{adj}(V)}{\det(V)};\;\;\text{adj}(V)= ( - 1)^{n + 1} \sum\limits_{k = 1}^n {c_k V^{k - 1} } ,
\label{eq:5}
\end{equation}
where the coefficients in the adjugate $\text{adj}(V)$ are
\begin{equation}
c_k  = \frac{1}{{n + 1}}\sum\limits_{j = 0}^n {e^{i\frac{{2\pi }}{{n + 1}}jk} \det (e^{ - i\frac{{2\pi }}{{n + 1}}j} I - V)}.
\label{eq:6}
\end{equation}
Furthermore, the determinant of any $n\times n$ matrix $B$ is
\begin{equation}
\det (B) = \sum\limits_{k_1 , \ldots ,k_n  = 1, \ldots ,1}^{n, \ldots ,n} {\!\!\!\!\!\!\!\!\!\varepsilon _{k_1 , \ldots ,k_n } \prod\limits_{q=1}^{n} {B_{q,k_q }} },
\label{eq:7}
\end{equation}
where $\varepsilon _{k_1 , \ldots ,k_n }$ is the Levi-Civita symbol \cite[]{RiLe}, given by
\begin{equation}
\varepsilon _{k_1 , \ldots ,k_n }  = \prod\limits_{a = 1,b = a + 1}^{n - 1,n} \!\!\!\!\!\!\!{{\mathop{\rm sgn}} (k_b  - k_a )}.
\label{eq:8}
\end{equation}
Thus, \Eq{7} and \Eq{8} show that we can indeed express \Eq{4} explicitly in terms of the \textit{elements} of $V$ and therefore also in terms of the $\bm{\lambda}$ of $A$ from \Eq{3} (see \App{B}).

As{\kern -0.7pt} our{\kern -0.7pt} main{\kern -0.7pt} example,{\kern -0.7pt} we{\kern -0.7pt} use{\kern -0.7pt} $f(A){\kern -0.5pt}={\kern -0.5pt}e^{tA}${\kern -0.7pt} for{\kern -0.7pt} scalar{\kern -0.7pt} $t$,{\kern -0.7pt} a{\kern -0.7pt} function{\kern -0.7pt} of{\kern -0.7pt} much{\kern -0.7pt} interest{\kern -0.7pt} \cite[]{MoV1,MoV2,LuRo,CuFZ,Curt},{\kern -0.7pt} given{\kern -0.7pt} here{\kern -0.7pt} by{\kern -0.7pt} \Eq{1}{\kern -0.7pt} as
\begin{equation}
e^{tA}  = \sum\limits_{k = 1}^n {A^{k - 1} \sum\limits_{\alpha = 1}^M {\sum\limits_{\beta = 1}^{\mu _{\alpha} } {(V^{ - 1} )_{k,c_{\alpha,\beta} } t^{\beta - 1} e^{\lambda _{\alpha} t} } } } ,
\label{eq:9}
\end{equation}
with $V^{-1}$ from \Eq{4} and $c_{\alpha,\beta}$ from \Eq{2}.  For $A\neq A(t)$, \Eq{9} solves
$\partial _t \mathbf{x}(t) = A\mathbf{x}(t)$ as mentioned earlier, but \Eq{9} is true in general, regardless of any variables in $A$ (dependencies would only affect which differential equations it solves).

We will elaborate on the many uses of \Eq{9} in \Sec{II}, but for now we list a few special cases that arise due to particular eigenvalue conditions.
\subsection{\label{sec:I.A}Special Case of $n$ Distinct Eigenvalues}
When $A$ has all $n$ eigenvalues different from each other $\lambda _1  \ne  \cdots  \ne \lambda _n$ so that $M=n$ and $
\bm{\mu}  = (1_1 , \ldots ,1_n )$, then the elements of the inverse confluent Vandermonde are
\begin{equation}
V_{a,b}^{ - 1}= \frac{{( - 1)^{n - a} e_{n - a} (\{ \bm{\lambda} \} \backslash \lambda _b )}}{{\prod\limits_{c = 1 \ne b}^n \!\!{(\lambda _b  - \lambda _c )} }},
\label{eq:10}
\end{equation}
where{\kern -0.8pt} the{\kern -0.8pt} notation{\kern -0.8pt} $\{ \bm{\lambda} \} \backslash \lambda _b${\kern -0.8pt} means{\kern -0.8pt} all{\kern -0.8pt} distinct{\kern -0.8pt} eigenvalues \textit{except}{\kern -0.8pt} $\lambda_b${\kern -0.8pt} (so{\kern -0.8pt} $\text{dim}(\{ \bm{\lambda} \} \backslash \lambda _b){\kern -0.8pt}={\kern -0.8pt}n{\kern -0.4pt}-{\kern -0.4pt}1$),{\kern -0.8pt} and{\kern -0.8pt} $e_j (\mathbf{x})${\kern -0.8pt} are{\kern -0.8pt} elementary symmetric polynomials of $m$ variables, given by
\begin{equation}
e_j (\mathbf{x}) \equiv \left\{ {\begin{array}{*{20}l}
   {1;} & {j = 0}  \\
   {\sum\limits_{1 \le k_1  <  \cdots  < k_j  \le m}\!\!\!\!\!\!\!\!\!\!\!\!\! {x_{k_1 }  \cdots x_{k_j } } ;} & {j \in 1, \ldots ,m}  \\
   {0;} & {j > m,}  \\
\end{array}} \right.
\label{eq:11}
\end{equation}
where $m\equiv\text{dim}(\mathbf{x})$.  An easier form to implement may be
\begin{equation}
e_j (\mathbf{x}) = \frac{{( - 1)^j }}{{m + 1}}\sum\limits_{k = 0}^m {e^{i\frac{{2\pi }}{{m + 1}}k(m - j)} \prod\limits_{l = 1}^m {(e^{ - i\frac{{2\pi }}{{m + 1}}k}  - x_l )} } ,
\label{eq:12}
\end{equation}
for $j \in 0, \ldots ,m$ and $m\equiv\text{dim}(\mathbf{x})$, derived in \App{D}.  See \App{C} for a symbolic-software-friendly version of \smash{$e_j (\mathbf{x}) $}. Using \Eq{10} in \Eq{9} gives the $n$-distinct case of \Eq{9} as
\begin{equation}
e^{tA}= \sum\limits_{k = 1}^n {A^{k - 1} \sum\limits_{\alpha  = 1}^n {\frac{{( - 1)^{n - k} e_{n - k} (\{ \bm{\lambda} \} \backslash \lambda _\alpha  )}}{{\prod\limits_{c = 1 \ne \alpha }^n {\!\!(\lambda _\alpha   - \lambda _c )} }}e^{\lambda _\alpha  t} } }.
\label{eq:13}
\end{equation}
\subsection{\label{sec:I.B}Special Case of $n$ Degenerate Eigenvalues}
When all $n$ eigenvalues of $A$ are the same $\lambda _1  =  \cdots  = \lambda _n$ so that $M=1$ and $
\bm{\mu}  = (n)$, then $V^{-1}$ has elements
\begin{equation}
\begin{array}{*{20}l}
   {V_{a,b}^{ - 1} } &\!\! { = \frac{{V_{a,b} ( - \lambda _1 )}}{{V_{a,a} V_{b,b} }}} &\!\! { = \left\{ {\begin{array}{*{20}l}
   {0;} & {b < a}  \\
   {\frac{{( - \lambda _1 )^{b - a} }}{{(b - a)!(a - 1)!}};} & {b \ge a.}  \\
\end{array}} \right.}  \\
\end{array}
\label{eq:14}
\end{equation}
Using \Eq{14} in \Eq{9} gives the $n$-degenerate case of \Eq{9} as
\begin{equation}
e^{tA}= e^{\lambda _1 t} \sum\limits_{k = 0}^{n - 1} {\frac{{(tA)^k }}{{k!}}\sum\limits_{\beta = 0}^{n - k - 1} {\frac{{( - \lambda _1 t)^\beta }}{{\beta!}}} }, 
\label{eq:15}
\end{equation}
where, whether the eigenvector matrix $U$ of $A$ is unitary or not, since $A=U\lambda_{1}IU^{-1}=\lambda_{1}I$, we get $e^{tA}\propto I$ as
\begin{equation}
\begin{array}{*{20}l}
   {e^{tA} } &\!\! { = \left({e^{\lambda _1 t} \sum\limits_{k = 0}^{n - 1} {\frac{{(\lambda _1 t)^k }}{{k!}}\sum\limits_{\beta = 0}^{n - k -1} {\frac{{( - \lambda _1 t)^\beta }}{{\beta!}}} }}\right)I=e^{\lambda _1 t}I, }  \\
\end{array}
\label{eq:16}
\end{equation}
and the unique eigenvalue of $e^{tA}$ is $e^{\lambda_1 t}$. This also yields the identity \smash{$\sum\nolimits_{a = 0}^{m} {\frac{{z^a }}{{a!}}\sum\nolimits_{b = 0}^{m - a} {\frac{{( - z)^b }}{{b!}}} }=1$}, for $m\in0,\ldots,\infty$.
\subsection{\label{sec:I.C}Intermediate Cases}
In general, \Eq{9} with input from \Eq{4} constructed using \Eq{3} solves all cases of $e^{tA}$, for any multiplicity structure.

For a given $n$, the number of different cases of multiplicities is given by the partition function $p(n)$, the number of ways of writing any integer $n$ as a sum of positive integers ignoring summand order, given by Euler's recursion formula \cite[]{Eule,Skie} (modified here to accept all $n$), 
\begin{equation}
p(n) = \delta _{n,0}  + \sum\limits_{k = 1}^n {( - 1)^{k + 1} [ p(q_{n,k}^ +  ) + p(q_{n,k}^ -  )] },
\label{eq:17}
\end{equation}
where $q_{n,k}^ \pm   \equiv n - \frac{1}{2}k(3k \pm 1)$, and by its definition $p(n<0)=0$, and by convention $p(0)\equiv 1$.

For example, for $n=4$, $\bm{\mu}$ can be $\bm{\mu}{\kern -0.7pt}={\kern -0.7pt}(1,1,1,1)$, $\bm{\mu}{\kern -0.7pt}={\kern -0.7pt}(1,1,2)$, $\bm{\mu}{\kern -0.7pt}={\kern -0.7pt}(2,2)$, $\bm{\mu}{\kern -0.7pt}={\kern -0.7pt}(1,3)$, or $\bm{\mu}{\kern -0.7pt}={\kern -0.7pt}(4)$, so that $\bm{\lambda}'{\kern -0.7pt}={\kern -0.7pt}(\lambda _1 ,\lambda _2 ,\lambda _3 ,\lambda _4)$, $\bm{\lambda}'{\kern -0.7pt}={\kern -0.7pt}(\lambda _1 ,\lambda _2 ,\lambda _3 ,\lambda _3 )$, $\bm{\lambda}'{\kern -0.7pt}={\kern -0.7pt}(\lambda _1 ,\lambda _1 ,\lambda _2 ,\lambda _2 )$, $\bm{\lambda}'{\kern -0.9pt}={\kern -0.9pt}(\lambda _1 ,\lambda _2 ,\lambda _2 ,\lambda _2 )$, or $\bm{\lambda}'{\kern -0.9pt}={\kern -0.9pt}(\lambda _1 ,\lambda _1 ,\lambda _1 ,\lambda _1 )$, respectively, giving $p(4){\kern -1.1pt}={\kern -1.1pt}5$ cases, where $\bm{\lambda}'{\kern -1.1pt}\equiv{\kern -1.1pt}({\lambda'}_{\!1},\ldots,{\lambda'}_{\!n})$ is the set of \textit{all} eigenvalues including repetitions while the unprimed $\lambda_k$ are distinct eigenvalues.  Since  \Sec{I.A} and \Sec{I.B} give special forms for the two extreme cases, there are $p(n)-2$ cases left for each $n$ where further simple forms may exist.  However, since $p(n)-2$ grows quickly with $n$, it may be generally easier to use \Eq{9} with \Eq{4}.
\section{\label{sec:II}Important Applications}
\subsection{\label{sec:II.A}Time-Evolution Operations}
A central task of quantum mechanics is to solve an equation of motion, such as the Schr{\"o}dinger equation $\partial _t |\psi (t)\rangle  {\kern -0.8pt}= {\kern -2pt} -\! \frac{i}{\hbar }H|\psi (t)\rangle$ or the von Neumann equation \smash{$\partial _t \rho (t) =\! -[\frac{i}{\hbar }H,\rho (t)]$}, for the quantum state of a system as function of time, given some Hamiltonian operator $H$ for the energy. For the von Neumann equation, states are represented as \textit{density operators} $\rho (t) \equiv \rho (t,t_0 ) \equiv |\psi (t)\rangle \langle \psi (t)|$ with the most general initial state being a mixture $\rho (t_0 ) = \sum\nolimits_j {p_j |\psi _j (t_0 )\rangle \langle \psi _j (t_0 )|} $, where $\sum\nolimits_j {p_j }  = 1$ such that $p_j  \in [0,1]$, and $|\psi _j (t_0 )\rangle$ are different pure states with $\langle \psi _j (t_0 )|\psi _j (t_0 )\rangle  = 1$, and $[A,B] \equiv AB - BA$.

Generally, the solution for all closed and some open systems has the form $\rho(t,t_{0})=U(t,t_{0})\rho(t_{0})U^{\dag}(t,t_{0})$, or if the initial state is pure it may also be represented as $|\psi(t )\rangle=U(t,t_{0})|\psi(t_0 )\rangle$, where in both cases $U(t,t_{0})$ is the unitary \textit{time-evolution operator}.  Another popular similar equation of motion is the Heisenberg equation, where time-dependence is grouped with observables rather than the state, but in \textit{both} Heisenberg and von Neumann pictures, all essential dynamics are contained in the Schr{\"o}dinger equation for the time-evolution operator (SETEO) $\partial _t U(t,t_{0})  {\kern -0.8pt}= {\kern -1pt} - \frac{i}{\hbar }HU(t,t_{0})$.

More generally, all open-system time evolution can be expressed as quantum operations \cite[]{HKr1,HKr2,Choi,Kra1} of the form \smash{$\rho(t,t_{0})\!=\!\sum\nolimits_{k}E_{k}(t,t_{0})\rho(t_{0})E_{k}^{\dag}(t,t_{0})$}, where \smash{$E_{k}(t,t_{0})$} are time-dependent{\kern -1pt} Kraus{\kern -1pt} operators{\kern -1pt} in{\kern -1pt} the{\kern -1pt} Hilbert{\kern -1pt} space{\kern -1pt} of{\kern -1pt} the{\kern -1pt} system,{\kern -1pt} satisfying{\kern 0.4pt} \smash{$\sum\nolimits_{k}\!E_{k}^{{\kern 0.1pt}\dag}(t,t_{0})E_{k}(t,t_{0}){\kern -0.2pt}={\kern -0.2pt}I$}.{\kern -1.8pt} In{\kern -1pt} this case, it is possible to find a larger effective joint closed system that evolves with a joint-system unitary of some effective joint-system Hermitian Hamiltonian \cite[]{HedD}.  Therefore, we will focus only on unitary evolution operators here, since they can model open systems as just described.

Thus, we now show how to apply the exponential operator formulas of this paper to the three types of unitary time-evolution operators \cite[]{Saku}, followed by examples.
\subsubsection{\label{sec:II.A.1}Time-Independent Hamiltonian}
If $H \ne H(t)$, then the solution to the SETEO,
\begin{equation}
U(t,t_0 ) = e^{ - \frac{i}{\hbar }(t - t_0 )H},
\label{eq:18}
\end{equation}
is given explicitly by $e^{tA}$ from \Eq{9} with $t \to  - \frac{i}{\hbar }(t - t_0 )$ and $A \to H$ with $n=\text{dim}(H)$, yielding (using $\Delta t\equiv t-t_{0}$),
\begin{equation}
U(t,t_0 ) = \sum\limits_{k = 1}^n {H^{k - 1}\! \sum\limits_{\alpha  = 1}^M {\sum\limits_{\beta  = 1}^{\mu _\alpha  } {V^{ - 1}_{k,c_{\alpha ,\beta } } ( - {\textstyle{i \over \hbar }}\Delta t)^{\beta  - 1} e^{ - \frac{i}{\hbar }\lambda _{\alpha}\Delta t  } } } },
\label{eq:19}
\end{equation}
where $\{\lambda_{\alpha}\}$ are now the distinct eigenvalues of $H$ (which are often known), and $M,\bm{\mu},c_{\alpha,\beta},V$ are as given in \Eqs{1}{4}.
\subsubsection{\label{sec:II.A.2}Time-Dependent, Time-Commuting Hamiltonian}
If $H = H(t)$ and $[H(t),H(t')]=0$ for $t\neq t'$, then 
\begin{equation}
U(t,t_0 ) = e^{ - \frac{i}{\hbar }\int_{t_0 }^t {H(t')dt'} }
\label{eq:20}
\end{equation}
is given by $e^{tA}$ from \Eq{9} with $t \to  - \frac{i}{\hbar }$ and $A \to \widetilde{H\rule{0pt}{7.5pt}} \equiv$ \smash{$\widetilde{H\rule{0pt}{7.5pt}}(t)\equiv\widetilde{H\rule{0pt}{7.5pt}}(t,t_{0})\equiv \int_{t_0 }^{\shiftmath{-0.6pt}{t}} {H(t')dt'}$}\rule{0pt}{8.2pt} with $n=\text{dim}(H)$ so that
\begin{equation}
U(t,t_0 ) = \sum\limits_{k = 1}^n {[\widetilde{H}(t)]^{k - 1} \!\sum\limits_{\alpha  = 1}^{M(t)} {\sum\limits_{\beta  = 1}^{\mu _\alpha^{(t)}} {\! V_{k,c_{\alpha ,\beta } }^{ - 1(t)} ( - {\textstyle{i \over \hbar }})^{\beta  - 1} e^{ - \frac{i}{\hbar }\widetilde{\lambda}_\alpha ^{(t)} } } } },
\label{eq:21}
\end{equation}
where we also put $\lambda _{\shiftmath{0.7pt}{\alpha}} \!\to\! \widetilde{\lambda}_{\shiftmath{0.7pt}{\alpha}} \!\equiv\! \widetilde{\lambda}_{\shiftmath{0.7pt}{\alpha}}  (t,t_0 )\! \equiv\! \widetilde{\lambda}_{\shiftmath{0.7pt}{\alpha}} ^{(t)} $ and \smash{$V^{ - 1}  \to $} \smash{$V^{ - 1(t)}  \!\equiv\! V^{ - 1} (\widetilde{\bm{\lambda}} ^{\shiftmath{-0.4pt}{(t)}} )$}, where \rule{0pt}{8.5pt}\smash{$\widetilde{\bm{\lambda}}^{\shiftmath{-0.4pt}{(t)}}  \!\equiv\! (\widetilde{\lambda}_1^{\shiftmath{0.4pt}{(t)}} , \ldots ,\widetilde{\lambda}_{M(t)}^{\shiftmath{0.4pt}{(t)}} )$} are the distinct eigenvalues of \smash{$\widetilde{H\rule{0pt}{7.0pt}}$} and can vary in time as can their{\kern -0.7pt} multiplicities{\kern -0.7pt} \smash{$\bm{\mu} ^{\shiftmath{-0.4pt}{(t)}}\! \equiv\! (\mu _1^{\shiftmath{0.5pt}{(t)}} {\kern -1pt}, \ldots ,\mu _{M(t)}^{\shiftmath{0.5pt}{(t)}} )$}{\kern -0.7pt} and{\kern -0.7pt} their{\kern -0.7pt} total number \smash{$M(t)$}, where arguments are in superscripts to avoid confusing them as factors.  Thus, at each time, the multiplicity structure of the eigenvalues of \smash{$\widetilde{H\rule{0pt}{7.5pt}}$} must be redetermined, and the structure of $V$ adapts accordingly.
\subsubsection{\label{sec:II.A.3}Time-Dependent, Non-Time-Commuting Hamiltonian}
If $H = H(t)$ and $[H(t),H(t')]\neq 0$, then the Suzuki-Trotter formula \cite[]{Suzu} for time-ordered exponentials is
\begin{equation}
\begin{array}{*{20}l}
   {U(t,t_0 )} &\!\! {= \mathop {\lim }\limits_{N \to \infty } \prod\limits^ \leftarrow  {}_{r = 1}^N e^{\frac{{ - i}}{\hbar }H(t_r )\Delta t_N }}  \\
   {} &\!\! { \approx e^{\frac{{ - i}}{\hbar }H(t_N )\Delta t_N } \! \cdots e^{\frac{{ - i}}{\hbar }H(t_2 )\Delta t_N } e^{\frac{{ - i}}{\hbar }H(t_1 )\Delta t_N } }  \\
   {} &\!\! {\approx U'(t_{N},t_{N-1})\cdots U'(t_{2},t_{1})U'(t_{1},t_{0}),} \\
\end{array}
\label{eq:22}
\end{equation}
where the product's arrow means it grows leftwards, and 
\begin{equation}
\Delta t_N  \!\equiv\! \Delta t_N (t) \! \equiv \!\frac{{t - t_0 }}{N}\;\;\,\text{and}\;\;\,t_r \! \equiv\! t_r (t) \!\equiv\! t_0  + r\Delta t_N ,
\label{eq:23}
\end{equation}
where $N\geq 1$ is as large as it needs to be for convergence, and \smash{$U'(t_{r},t_{r-1})\equiv e^{ - i(t_{r}-t_{r-1})H(t_r )/\hbar }$}.

Here, each factor of $e^{ - iH(t_r )\Delta t_N /\hbar }$ in \Eq{22} is given by \smash{$e^{tA}$} from \Eq{9} with \smash{$t \to  - \frac{i}{\hbar }\Delta t_N$} and \smash{$A \to H(t_r )$}, so
\begin{equation}
\begin{array}{*{20}l}
   {U(t,t_0 )\approx } &\!\!\! {\prod\limits_{r = 1}^{\scriptstyle  \leftarrow  \hfill \atop 
  \scriptstyle N \hfill} {\left(\! {\rule{0pt}{14pt}} \right.\sum\limits_{k = 1}^n {[H(t_r )]^{k - 1} } } }  \\
   {} &\!\!\! { \times\! \sum\limits_{\alpha  = 1}^{M(t_r )} {\sum\limits_{\beta  = 1}^{\mu _\alpha ^{(t_r )} } {V_{k,c_{\alpha ,\beta } }^{ - 1(t_r )} (\frac{{ - i}}{\hbar }\Delta t_N )^{\beta  - 1} e^{\frac{{ - i}}{\hbar }\lambda _\alpha ^{(t_r )} \Delta t_N } } } \left. {\rule{0pt}{14pt}}\!\!\!\! \right)\!,}  \\
\end{array}
\label{eq:24}
\end{equation}
where \smash{$\lambda _\alpha \!\to\! \lambda _\alpha  (t_r ) {\kern -0.6pt}\equiv{\kern -0.6pt} \lambda _\alpha ^{(t_r )}$}, \smash{$V^{ - 1}  \!\to\! V^{ - 1(t_r )}  \equiv V^{ - 1} (\bm{\lambda} ^{\shiftmath{-0.5pt}{(t_r )}} )$}, and{\kern -1.5pt} \smash{$\bm{\lambda}^{{\kern -0.9pt}(t_r )} {\kern -0.7pt} \!\equiv\!{\kern -0.5pt} ({\kern -0.5pt}\lambda _{{\kern -0.5pt}1}^{{\kern 1.2pt}\shiftmath{-0.4pt}{(t_r )}} {\kern -0.5pt}\!, \ldots ,\!\lambda _{{\kern -0.5pt}M{\kern -0.7pt}(t_r )}^{{\kern 0.5pt}\shiftmath{0.3pt}{(t_r )}} )$}{\kern -1pt} are{\kern -1pt} the{\kern -1pt} $M(t_r )${\kern -1pt} distinct{\kern -1pt} eigenvalues{\kern 1.9pt} of{\kern 1.9pt} $H(t_r )$,{\kern 1.9pt} where{\kern 1.9pt} \smash{$\bm{\mu} ^{{\kern 0.7pt}\shiftmath{-1.2pt}{(t_r )}} \! \equiv \!(\mu _{{\kern -0.5pt}1}^{{\kern 1.2pt}\shiftmath{0.3pt}{(t_r )}}\! , \ldots ,\mu _{{\kern -0.5pt}M{\kern -0.5pt}(t_r )}^{\shiftmath{0.3pt}{(t_r )}} {\kern -0.5pt})$}.  Thus, we{\kern -0.7pt} can{\kern -0.7pt} get{\kern -0.7pt} a{\kern -0.7pt} \textit{symbolic}{\kern -0.7pt} expression{\kern -0.7pt} for{\kern -0.7pt} $U(t,t_0 )$ to \textit{any} degree of approximation. Convergence is discussed in \Sec{III}.

This same procedure applies to general time-ordered exponentials \smash{$\tau \{ e^{\shiftmath{0.2pt}{\int_{t_0 }^t {\kern -0.5pt}{A(t')dt'} }} \}$} where $\tau \{  \cdot \}$ is the time ordering operator, with \smash{$\frac{-i}{\hbar} \to 1$} and \smash{$H(t_r ) \to A(t_r )$} in \Eq{24}.  Time-ordered exponentials arise in many areas, such as in linearized quantum Langevin equations \cite[]{GaCo,GaZo}.  See \cite[]{GLTJ} for an intriguing method to get explicit time-ordered exponentials with path integrals.
\subsubsection{\label{sec:II.A.4}Example: Exact Solution of All Single-Qubit Schr{\"o}dinger and von Neumann Equations}
For a single qubit, $n=2$, given a Hermitian Hamiltonian $H$, there are two multiplicity cases for eigenvalues of $H$; $\lambda_1 \neq \lambda_2$ and $\lambda_1 =\lambda_2$.  If the time evolution of the system is unitary then there are three main cases.

\textit{Case 1:} If $H \ne H(t)$, then in subcases where $\lambda_1 \neq \lambda_2$,
\begin{equation}
\begin{array}{*{20}l}
   {U(t,t_0 )=} &\!\! {\frac{1}{{\lambda _1  - \lambda _2 }}\left[ {(\lambda _1 e^{\frac{{ - i}}{\hbar }(t - t_0 )\lambda _2 }  - \lambda _2 e^{\frac{{ - i}}{\hbar }(t - t_0 )\lambda _1 } )I} \right.}  \\
   {} &\!\! {\left. { + (e^{\frac{{ - i}}{\hbar }(t - t_0 )\lambda _1 }  - e^{\frac{{ - i}}{\hbar }(t - t_0 )\lambda _2 } )H} \right],}  \\
\end{array}
\label{eq:25}
\end{equation}
and in subcases where $\lambda_1 =\lambda_2$,
\begin{equation}
U(t,t_0 ) = e^{\frac{{ - i}}{\hbar }(t - t_0 )H_{1,1} } I.
\label{eq:26}
\end{equation}

In both of the above subcases, since $n<5$, we can get the eigenvalues explicitly in terms of elements of $H$ as
\begin{equation}
\begin{array}{*{20}l}
   {\lambda _{\scriptstyle 1 \hfill \atop 
  \scriptstyle 2 \hfill} } &\!\! { = \frac{{\text{tr}(H)}}{2} \pm \sqrt {(\frac{{\text{tr}(H)}}{2})^2  - \det (H)} }  \\
   {} &\!\! { = \frac{{H_{1,1}  + H_{2,2}  \pm \sqrt {(H_{1,1}  - H_{2,2} )^2  + 4|H_{1,2} |^2 } }}{2},}  \\
\end{array}
\label{eq:27}
\end{equation}
in which case the condition $\lambda_1 \neq \lambda_2$ becomes $H \ne H_{1,1} I$, and the condition $\lambda_1 = \lambda_2$ becomes $H = H_{1,1} I$.  Thus, putting \Eq{27} into \Eq{25} gives
\begin{equation}
U(t,t_0 ) = e^{ - {\textstyle{i \over \hbar }}(t - t_0 ){\textstyle{{H_{1,1}  + H_{2,2} } \over 2}}} \!\left( {\begin{array}{*{20}c}
   a & b  \\
   { - b^* } & {a^* }  \\
\end{array}} \right)\!,
\label{eq:28}
\end{equation}
where $a$ and $b$ automatically obey $|a|^2 \!+\!|b|^2 \!=\!1$, given by 
\begin{equation}
\begin{array}{*{20}l}
   a &\!\! { \equiv \cos (\omega _H \Delta t) - i\frac{{\Delta _H }}{{\hbar{\kern 0.5pt} \omega _H +\delta_{\omega_{H},0}}}\sin (\omega _H \Delta t)}  \\
   b &\!\! { \equiv  - i\frac{{H_{1,2} }}{{\hbar{\kern 0.5pt} \omega _H +\delta_{\omega_{H},0}}}\sin (\omega _H \Delta t),}  \\
\end{array}
\label{eq:29}
\end{equation}
with $\Delta t\equiv t-t_{0}$, and using abbreviations
\begin{equation}
\omega _H  \equiv {\textstyle{1 \over \hbar }}\sqrt {|\Delta _H |^2  + |H_{1,2} |^2 } \;\;\;\text{and}\;\;\;\Delta _H  \equiv {\textstyle{{H_{1,1}  - H_{2,2} } \over 2}},
\label{eq:30}
\end{equation}
and the $\delta_{\omega_{H},0}$ in \Eq{29} lets \Eq{28} hold for both multiplicities. Therefore, \Eq{28} \textit{exactly} solves all Schr{\"o}dinger and von Neumann equations for constant $H$. Thus, \Eq{28} lets us design any single-qubit unitary gate we want. 

As an elementary example of gate design, to get a Hadamard gate \smash{$U_H  = \frac{1}{{\sqrt 2 }}(_{1\; - 1}^{1\;\;\;\,1} )$}, setting $U(t,t_0 )=U_H$ in \Eq{28} shows that, up to global phase, this can be exactly achieved with \smash{$H  = \hbar(_{1\; - 1}^{1\;\;\;\,1} )$} at times \smash{$t = t_0  + {\textstyle{1  \over {\sqrt 2 }}}({\textstyle{\pi  \over 2}} + 2\pi k)$} for $k \in \{ 0,1,2, \ldots \}$.  Similarly, a phase gate \smash{$U_\phi  = (_{0\;\, e^{i\phi}}^{{\kern 0.5pt}1\;\;\;0} )$} can be exactly achieved, up to global phase, by \smash{$H = \hbar(_{0\; -\Delta}^{\Delta\;\;\;0} )$} with $\Delta>0$ at times \smash{$t = t_0  + {\textstyle{1  \over \Delta }}({\textstyle{\phi  \over 2}} + 2\pi k)$}.  Together, these two kinds of gates can be combined to produce arbitrary single-qubit gates.

\textit{Case 2:} If $H = H(t)$ and $[H(t),H(t')]=0$ for $t\neq t'$, then $U(t,t_0)$ is given by \Eqs{25}{30} with $(t - t_0 )\to 1$, $H\to \widetilde{H\rule{0pt}{7pt}}$, and \smash{$\lambda _{\alpha} \to \widetilde{\lambda}_{\alpha}^{(t)}$} where \smash{$\widetilde{H\rule{0pt}{7pt}}\equiv\widetilde{H\rule{0pt}{7pt}}(t,t_{0})\equiv \int_{t_0 }^t {\!H(t')dt'}$} with distinct eigenvalues \smash{$\widetilde{\lambda}_{\alpha}^{(t)}$}, which applied to \Eq{28}, yields
\begin{equation}
U(t,t_0 ) = e^{ - {\textstyle{i \over \hbar }}{\textstyle{{\widetilde{H}_{1,1}  + \widetilde{H}_{2,2} } \over 2}}}\! \left( {\begin{array}{*{20}c}
   {\widetilde{a}} & {\widetilde{b}}  \\
   { - \widetilde{b}^* } & {\widetilde{a}^* }  \\
\end{array}} \right)\!,
\label{eq:31}
\end{equation}
where \Eq{29} is likewise modified to give
\begin{equation}
\begin{array}{*{20}l}
   {\widetilde{a}} &\!\! {\equiv \cos (\omega _{\widetilde{H}} ) - i{\textstyle{{\Delta _{\widetilde{H}} } \over {\hbar \omega _{\widetilde{H}}  + \delta _{\omega _{\widetilde{H}} ,0} }}}\sin (\omega _{\widetilde{H}} )}  \\
   {\widetilde{b}} &\!\! {\equiv  - i{\textstyle{{\widetilde{H}_{1,2} } \over {\hbar \omega _{\widetilde{H}}  + \delta _{\omega _{\widetilde{H}} ,0} }}}\sin (\omega _{\widetilde{H}} ),}  \\
\end{array}
\label{eq:32}
\end{equation}
where $\omega _{\widetilde{H}}  \equiv {\textstyle{1 \over \hbar }}\sqrt {|\Delta _{\widetilde{H}} |^2  + |\widetilde{H}_{1,2} |^2 }$ and $\Delta _{\widetilde{H}}  \equiv {\textstyle{{\widetilde{H}_{1,1}  - \widetilde{H}_{2,2} } \over 2}}$.

\textit{Case 3:} If $H = H(t)$ and $[H(t),H(t')]\neq 0$, then in the $H \ne H_{1,1} I$ subcase, $U(t,t_0)$ is given by \Eq{22}, where each factor of the form $e^{ - iH(t_r )\Delta t_N /\hbar }$ is given by \Eqs{28}{30} with $(t - t_0 )\to \Delta t_N$, $H\to H (t_r )$, where $\Delta t_N$ and $t_r $ are from \Eq{23}, and \smash{$\lambda_{\alpha}\to\lambda_{\alpha}^{(t_r )}$} are distinct eigenvalues of $H (t_r )$, all of which, when put into \Eq{28} and then \Eq{22}, yields
\begin{equation}
U(t,t_0 ) \approx \prod\limits_{r = 1}^{\scriptstyle  \leftarrow  \hfill \atop 
  \scriptstyle N \hfill} {e^{ - {\textstyle{i \over \hbar }}\Delta t_{N}{\textstyle{{H_{1,1}^{(t_r )}  + H_{2,2}^{(t_r )} } \over 2}}} \!\left( {\begin{array}{*{20}c}
   {a_r } & {b_r }  \\
   { - b_r ^* } & {a_r ^* }  \\
\end{array}} \right)\!,}
\label{eq:33}
\end{equation}
where \Eq{29} is likewise modified to become
\begin{equation}
\begin{array}{*{20}l}
   {a_r } &\!\! { \equiv \cos (\omega _{H(t_r )} \Delta t_{N}) - i{\textstyle{{\Delta _{H(t_r )} \sin (\omega _{H(t_r )} \Delta t_{N})} \over {\hbar \omega _{H(t_r )}  + \delta _{\omega _{H(t_r )},0 } }}}}  \\
   {b_r } &\!\! { \equiv  - i{\textstyle{{H_{1,2}^{(t_r )}\sin (\omega _{H(t_r )} \Delta t_{N}) } \over {\hbar \omega _{H(t_r )}  + \delta _{\omega _{H(t_r )},0 } }}},}  \\
\end{array}
\label{eq:34}
\end{equation}
\mbox{s.t.{\kern 3pt}\rule{0pt}{9.5pt}\smash{$\omega _{H(t_r )} \! \equiv\! {\textstyle{1 \over \hbar }}\sqrt {|\Delta _{H(t_r )} |^2  \!+\! |H_{1,2}^{(t_r )} |^2 } $}, \smash{$\Delta _{H(t_r )}  \!\equiv\! {\textstyle{{H_{1,1}^{(t_r )} \! -\! H_{2,2}^{(t_r )} } \over 2}}$}.} For\hsp{0.7} details\hsp{0.7} on\hsp{0.7} the\hsp{-0.7} convergence\hsp{-0.7} of\hsp{-0.7} \Eq{33},\hsp{-0.7} see\hsp{-0.7} \Sec{III}.

When \rule{0pt}{10pt}$H(t) = H_{1,1}(t) I$, the Hamiltonian is automatically time-commuting, and so even though we could use Case 3, an exact answer is assured more simply by using Case 2. For $n$-level systems, $U(t,t_{0})$ has a general hyperspherical form analogous to \Eq{28}, as shown in \cite[]{HedU}.

\begin{figure}[H]
\centering
\includegraphics[width=0.99\linewidth]{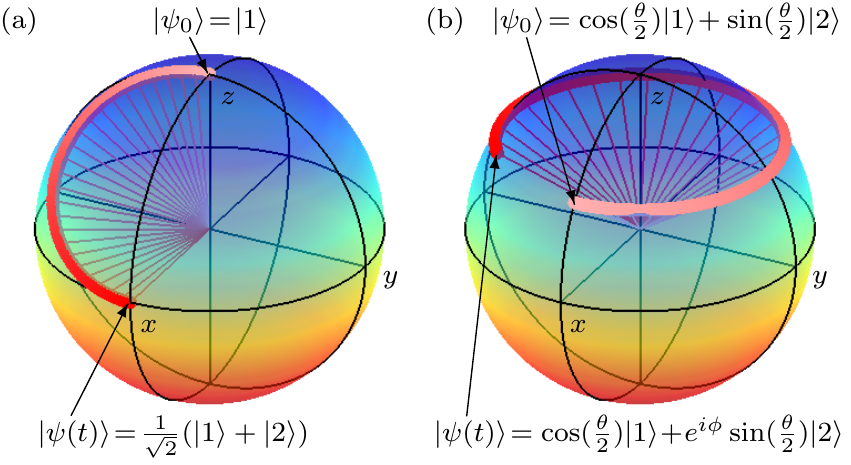}%
\caption[]{(color online) Exact evolution $|\psi(t)\rangle=U(t,t_{0})|\psi_{0}\rangle$ of a pure qubit with a constant Hamiltonian $H$ represented on the Bloch sphere, with the path colored to show the passage of time, with red as the final time. (a) shows the effects of \smash{$H  = \hbar(_{1\; - 1}^{1\;\;\;\,1} )$} from $t_0$ to \smash{$t = t_0  + {\textstyle{1  \over {\sqrt 2 }}}{\textstyle{\pi \over 2}}$} at which time the Hadamard gate's action is achieved, changing initial state $|\psi_0\rangle\!=\!|1\rangle$ into $|\psi(t)\rangle\!=\!\frac{1}{\sqrt{2}}(|1\rangle+|2\rangle)$. (b) shows the effects of \smash{$H  = \hbar(_{0\; -\Delta}^{\Delta\;\;\;0} )$} with $\Delta>0$ from $t_0$ to \smash{$t = t_0  + {\textstyle{1 \over \Delta }}{\textstyle{\phi  \over 2}}$} where $\phi=0.85(2\pi)$ at which time the phase gate's action is achieved, changing initial state \smash{$|\psi_0\rangle=\cos(\frac{\theta}{2})|1\rangle+\sin(\frac{\theta}{2})|2\rangle$} where $\theta=0.31\pi$ into \smash{$|\psi(t)\rangle=\cos(\frac{\theta}{2})|1\rangle+e^{i\phi}\sin(\frac{\theta}{2})|2\rangle$}. Bloch coordinates are $(x,y,z) \!\equiv \!(2x_1 x_2 ,2x_1 x_3 ,2x_1^2 {\kern -0.5pt} - {\kern -0.5pt}1)$,{\kern -0.5pt} where{\kern -0.5pt} $(x_1 ,x_2 ,x_3 )\! \equiv\! (\langle 1|\psi '\rangle ,\text{Re}[\langle 2|\psi '\rangle ],\text{Im}[\langle 2|\psi '\rangle ])$,{\kern -1pt} with{\kern -1pt} $|\psi '\rangle \! \equiv\! e^{ - i\arg [\langle 1|\psi(t) \rangle ]} |\psi (t)\rangle$.}
\label{fig:1}
\end{figure}

\Figure{1} plots the time evolution of a pure state on its Bloch sphere \cite[]{Blch,NiCh}, for the Hadamard and phase gate.
\subsection{\label{sec:II.B}Baker-Campbell-Hausdorff (BCH) Formula}
For operators $A$ and $B$ and scalar $s$, where $A$, $B$, and $s$ can be time-dependent, the BCH formula \cite[]{Saku,NiCh} is
\begin{equation}
\begin{array}{*{20}l}
   {e^{sA} Be^{ - sA} } &\!\! { = \sum\limits_{k = 0}^\infty  {\frac{{s^k }}{{k!}}([A,)^k B(])^k } }  \\
   {} &\!\! { = B + s[A,B] + \frac{{s^2 }}{{2!}}[A,[A,B]] +  \cdots ,}  \\
\end{array}
\label{eq:35}
\end{equation}
where, for example, $([A,)^0 B(])^0  = B$, and $([A,)^1 B(])^1  = [A,B]$, and $([A,)^2 B(])^2  = [A,[A,B]]$, etc.  The nested commutators in \Eq{35} are a common source of difficulty.

However, with our results, it is \textit{always} possible to express this quantity exactly as a product of three matrices, $e^{sA}$, $B$, and $e^{-sA}$.  In particular, in all cases of time-dependence, the explicit forms of $e^{\pm sA}$ are given by $e^{tA}$ from \Eq{9} with $t \to  \pm s$, which leads to the \textit{finite} sum,
\begin{equation}
\begin{array}{*{20}l}
   {e^{sA} Be^{ - sA}  {\kern -0.5pt}= } &\!\!{\kern -0.5pt} {\sum\limits_{k = 1}^n {\!\!A^{k - 1} B }\! \sum\limits_{\alpha  = 1}^M {} \!\sum\limits_{\beta  = 1}^{\mu _\alpha  } {}\!\! V_{k,c_{\alpha ,\beta } }^{ - 1} s^{\beta  - 1} e^{\lambda _\alpha  s} }  \\
   {} &\!\!{\kern -0.5pt} { \times\!\!\sum\limits_{k' = 1}^n {\!\!\!A^{k' - 1} } \!\!\sum\limits_{\alpha ' = 1}^M {}\! \sum\limits_{\beta ' = 1}^{\mu _{\alpha '} } {} \!\!V_{k',c_{\alpha ',\beta '} }^{ - 1} \!( - s)^{\beta ' - 1} e^{ - \lambda _{\alpha '} s}, }  \\
\end{array}
\label{eq:36}
\end{equation}
where $A$ is allowed to be time-dependent, as are $s$ and $B$, $c_{\alpha,\beta}$ is as defined for \Eq{2}, $n=\text{dim}(A)$, and $\lambda_{\alpha}$ are the distinct eigenvalues of $A$, all as in \Sec{I}.  Thus, the computation of all the commutators can \textit{always} be avoided \textit{and} we \textit{always} get an explicit solution, provided that we know the eigenvalues of $A$.

Note that the quantity \smash{$\tau{\kern -0.5pt} \{ e^{\int_{t_0 }^t {A(t')dt'} }\! \}B[\tau{\kern -0.5pt} \{ e^{\int_{t_0 }^t {A(t')dt'} }\! \}]^{\dag}\!$}, with time ordering operator $\tau \{  \cdot \}$, is \textit{not} solved by the BCH formula.  However, in this case we can still solve this as a product of three operators, where \smash{$\tau \{ e^{\shiftmath{0pt}{\int_{t_0 }^t {A(t')dt'}}} \}$} is approximated as an explicit function to any convergence for $N$ exponential operators, as in \Sec{II.A.3}, with the same convergence problems that all Suzuki-Trotter approximations have.  However, such difficulties may be able to be avoided completely using effective joint systems, as mentioned in \Sec{III}.
\subsection{\label{sec:II.C}Alternative Forms and General Examples}
The version of $V^{-1}$ in \Eq{4} can be expressed (using parenthetical superscripts as arguments) as
\begin{equation}
\begin{array}{*{20}l}
   {V^{ - 1}  =} &\!\! {\sum\nolimits_{l = 1}^n \!{q_l^{(V)} V^{l - 1} } ,}  \\
\end{array}
\label{eq:37}
\end{equation}
with $\mathbf{q}^{(V)}  \equiv (q_1^{(V)} , \ldots ,q_n^{(V)} )^{T}$ where
\begin{equation}
\begin{array}{*{20}l}
   {\mathbf{q}^{(V)}  \equiv W\mathbf{d}^{(V)};} & {\text{so}} & {q_l^{(V)}  = \sum\nolimits_{m = 1}^{n + 1} \!{W_{l,m} } d_m^{(V)}  ,}  \\
\end{array}
\label{eq:38}
\end{equation}
where \smash{$\mathbf{d}^{(V)}  \equiv (d_1^{(V)} , \ldots ,d_{n + 1}^{(V)} )^T $} and $W$ is an $n\times (n+1)$ matrix, each with elements\\
\vspace{-14pt}
\begin{equation}
\begin{array}{*{20}l}
   {W_{l,m} } &\!\! { \equiv \frac{1}{{\sqrt {n + 1} }}e^{i\frac{{2\pi }}{{n + 1}}l(m - 1)}}  \\
   {d_{m}^{(V)} } &\!\! { \equiv \frac{{( - 1)^{n + 1} }}{{\det (V)\sqrt {n + 1} }}\det (e^{ - i\frac{{2\pi }}{{n + 1}}(m - 1)} I - V),} \\
\end{array}
\label{eq:39}
\end{equation}
for \smash{$l  \in 1, \ldots ,n$}\rule{0pt}{20pt} and \smash{$m \in 1, \ldots ,n + 1$}, all of which allow \Eq{37} to compactly represent elements of \smash{$V^{-1}$} \mbox{as \smash{$V_{a,b}^{ - 1}  {\kern 0pt}={\kern -1pt}  \sum\nolimits_{\shiftmath{1pt}{l \!=\! 1}}^n {q_l^{{\kern 1pt}\shiftmath{0pt}{(V)}} (V^{\shiftmath{-1pt}{l \!-\! 1}} )_{a,b} } {\kern 0pt} ={\kern -2pt} \sum\nolimits_{\shiftmath{1pt}{l \!=\! 1}}^n {q_l^{{\kern 1pt}\shiftmath{0pt}{(V)}} \langle a|V^{\shiftmath{-1pt}{l \!-\! 1}} |b\rangle } {\kern 0pt} {\kern -1pt}=$}} \smash{$ \sum\nolimits_{\shiftmath{1pt}{l \!=\! 1}}^n {{\kern 0pt}q_l^{{\kern 1pt}\shiftmath{0pt}{(V)}} {\kern 0pt}\text{tr}(V^{\shiftmath{-1pt}{l \!-\! 1}} |b\rangle \langle a|} )$} where \smash{$\{ |1\rangle , \ldots ,|n\rangle \} $} is the complete orthonormal standard basis.

Then, using the indical register function \smash{$c_{\alpha,\beta}$} from \Eq{2}, we can express the general result from \Eq{1} as 
\begin{equation}
\begin{array}{*{20}l}
   {f(A)} &\!\! { =\! \sum\limits_{k = 1}^n {A^{k - 1}\! \sum\limits_{\alpha  = 1}^M {\sum\limits_{\beta  = 1}^{\mu _\alpha  } {V_{k,c_{\alpha ,\beta } }^{ - 1} \widetilde{f}_{c_{\alpha ,\beta } } } } } }  \\
   {} &\!\! { =\! \sum\limits_{k = 1}^n {A^{k - 1}\! \sum\limits_{\alpha  = 1}^M {\sum\limits_{\beta  = 1}^{\mu _\alpha  } {\sum\limits_{l = 1}^n {q_l^{(V)} (V^{l - 1} )_{k,c_{\alpha ,\beta } } } \widetilde{f}_{c_{\alpha ,\beta } } } } } }  \\
   {} &\!\! { =\! \sum\limits_{k = 1}^n {A^{k - 1} }\! \sum\limits_{l = 1}^n {q_l^{(V)} \!\sum\limits_{\alpha  = 1}^M {\sum\limits_{\beta  = 1}^{\mu _\alpha  } \!{(V^{l - 1} )_{k,c_{\alpha ,\beta } } \widetilde{f}_{c_{\alpha ,\beta } } } } }, }  \\
\end{array}
\label{eq:40}
\end{equation}
where \smash{$(V^{l - 1} )_{k,c_{\alpha ,\beta } } \equiv \langle k|V^{l - 1} |c_{\alpha ,\beta } \rangle = \text{tr}(V^{l - 1} |c_{\alpha ,\beta } \rangle \langle k|) $}, and \smash{$\widetilde{f}_{c_{\alpha ,\beta } } $} is as in \Eq{2}.  Alternatively, we could use \Eq{B.7} in \Eq{1} to get $f(A)$ explicitly in terms of the eigenvalues of $A$, but \Eq{40} has the advantage of being more compact while still expressing $f(A)$ in terms of $V$ rather than its inverse.  The benefit of expressing $f(A)$ with \Eq{1} and \Eq{4} rather than \Eq{40} is that \Eq{1} requires fewer auxiliary definitions, letting us see its dependence on $V$ more clearly.

Next, we give examples of $e^{tA}$ where $t$ is a scalar and $A$ is an $n\times n$ matrix with the same definition of distinct eigenvalue structure described in \Sec{I}. 

In $n=2$, for $\bm{\mu}=(1,1)$,
\begin{equation}
\begin{array}{*{20}l}
   {e^{tA}} &\!\! {=\frac{1}{{\lambda _1  - \lambda _2 }}\left[ {(\lambda _1 e^{\lambda _2 t}  \!-\! \lambda _2 e^{\lambda _1 t} )I \!-\! (e^{\lambda _2 t}  \!-\! e^{\lambda _1 t} )A} \right]\!,}  \\
\end{array}
\label{eq:41}
\end{equation}
and for $\bm{\mu}=(2)$,
\begin{equation}
e^{tA}  = e^{\lambda _1 t} [(1 - \lambda _1 t)I + tA] = e^{\lambda _1 t} I.
\label{eq:42}
\end{equation}

In $n=3$, for $\bm{\mu}=(1,1,1)$,
\begin{widetext}
\begin{equation}
\begin{array}{*{20}l}
   {e^{tA}  = } &\!\! {\frac{1}{{(\lambda _1  - \lambda _2 )(\lambda _1  - \lambda _3 )(\lambda _2  - \lambda _3 )}}\left\{ {[\lambda _1 \lambda _2 (\lambda _1  - \lambda _2 )e^{\lambda _3 t}  - \lambda _1 \lambda _3 (\lambda _1  - \lambda _3 )e^{\lambda _2 t}  + \lambda _2 \lambda _3 (\lambda _2  - \lambda _3 )e^{\lambda _1 t} ]I} \right.}  \\
   {} &\!\! {\left. { + [ - (\lambda _1 ^2  - \lambda _2 ^2 )e^{\lambda _3 t}  + (\lambda _1 ^2  - \lambda _3 ^2 )e^{\lambda _2 t}  - (\lambda _2 ^2  - \lambda _3 ^2 )e^{\lambda _1 t} ]A + [(\lambda _1  - \lambda _2 )e^{\lambda _3 t}  - (\lambda _1  - \lambda _3 )e^{\lambda _2 t}  + (\lambda _2  - \lambda _3 )e^{\lambda _1 t} ]A^2 } \right\},}  \\
\end{array}
\label{eq:43}
\end{equation}
for $\bm{\mu}=(1,2)$,
\begin{equation}
\begin{array}{*{20}l}
   {e^{tA}  = } &\!\! {\frac{1}{{(\lambda _1  - \lambda _2 )^2 }}\left( {\{ \lambda _1 [(\lambda _1  - 2\lambda _2 ) - \lambda _2 (\lambda _1  - \lambda _2 )t]e^{\lambda _2 t}  + \lambda _2 ^2 e^{\lambda _1 t} \} I} \right.}  \\
   {} &\!\! {\left. { + \{ [2\lambda _2  + (\lambda _1 ^2  - \lambda _2 ^2 )t]e^{\lambda _2 t}  - 2\lambda _2 e^{\lambda _1 t} \} A + \{  - [1 + (\lambda _1  - \lambda _2 )t]e^{\lambda _2 t}  + e^{\lambda _1 t} \} A^2 } \right),}  \\
\end{array}
\label{eq:44}
\end{equation}
\end{widetext}
and for $\bm{\mu}=(3)$,
\begin{equation}
\begin{array}{*{20}l}
   {e^{tA} } &\!\! { = e^{\lambda _1 t} [(1 - \lambda _1 t + \frac{1}{2}\lambda _1 ^2 t^2 )I + (1 - \lambda _1 t)tA + \frac{1}{2}t^2 A^2 ]} \\
   {} &\!\! {= e^{\lambda _1 t} I.} \\
\end{array}
\label{eq:45}
\end{equation}

As mentioned in \Sec{I}, these results also hold if $A$ is $t$-dependent, but such dependence will affect which differential equations are solved by $e^{tA(t)}$.

For excellent examples of the fully-distinct-eigenvalue exponentials, see \cite[]{CuFZ} which gives the single-mode spin operators using a Cayley-Hamilton expansion as we have here. However, for exponentials of \textit{multipartite} spin operators, there will generally be degeneracy in the eigenvalues, and in that case, the methods presented here are needed instead. Note that the examples in \Eqs{41}{45} are for dimensions that do not support multipartite systems, so for multipartite operators, use \Eq{9} or \Eq{40}.
\section{\label{sec:III}Conclusions}
We have provided a few new subtle improvements for achieving operator functions with Vandermonde methods.  While these improvements are very simple, they allow great simplifications for these kinds of problems, particularly with theoretical work.

The main simple improvements are the new formula for the \textit{matrix inverse} in \Eq{4} involving unit-complex exponentials, and the \textit{indical register function} $c_{\alpha,\beta}$ defined for \Eq{2}. The inverse function lets us express the inverse confluent Vandermonde $V^{-1}$ as a simple nonrecursive formula in terms of the confluent Vandermonde $V$ itself, without any constraints on the indices.  Meanwhile, $c_{\alpha,\beta}$ automatically takes care of mapping the $n\times n$ indices of $V$ to the correct matrix elements based on its multiplicity structure.  This allows all operator functions to be described as closed-form self-contained sums, as seen in the main example for exponential operators in \Eq{9}.

The motivation for studying the inverse confluent Vandermonde matrix in particular is that it appears in the Cayley-Hamilton theorem for general operator functions, allowing any analytic function of an operator to be expressed only in terms of that operator's eigenvalues and itself, \textit{without explicit appearance of its eigenvectors}, which would be considerably more complicated.

The special-case formulas for elements of $V^{-1}$ in \Eq{10} and \Eq{14} were found by observation and confirmed exactly through symbolic testing for $n\in 2,\ldots,8$.  The true usefulness of \Eq{4} really becomes apparent when we consider the \textit{intermediate} cases of multiplicity, the number of which rapidly increases with $n$, as explained in \Sec{I.C}.  In those cases, \Eq{4} provides a particularly simple form for $V^{-1}$, \textit{valid for all multiplicity structures}.

Since exponential operators are of particular importance in quantum mechanics, we focused on that for our main examples.  We showed how to get explicit forms of the time-evolution operator $U(t,t_{0})$ for the three main cases, with the non-time-commuting case being merely an approximation, but which can be achieved to any degree of accuracy, remaining symbolic in all cases.  

To verify how well the approximation of \Eq{24} works for a given $H(t)$, $N$, and $\Delta t\equiv t-t_0$, \Eq{24} can be tested in both sides of the SETEO $\partial _t U(t,t_{0})  {\kern -0.8pt}= {\kern -1pt} - \frac{i}{\hbar }H(t)U(t,t_{0})$, to see how well their difference approximates the zero matrix. It is important to note that a SETEO test requires that we retain full symbolic $t$-dependence in $\Delta t_N$ and $t_{r}$ from \Eq{23} to get the correct $\partial _t U(t,t_{0})$. This test worked well for small dimensions $n$ and orders $N$ but when that is impractical, we must rely on tests showing that $H(t')$ is approximately constant during $\Delta t_N (t')$ for all $t'\in[t_{0},t]$.  Furthermore, convergence and stability depend strongly on the magnitudes of the eigenvalues of $H(t)$, which should ideally be less than unity.

However, despite any convergence difficulties that may arise with \Eq{24}, the unitary dynamics of systems of non-time-commuting Hamiltonians may be able to be modeled as reductions of larger effective finite systems with constant or time-commuting Hamiltonians, and then the results of \Eq{19} or \Eq{21} can be used quite effectively.

In particular, we pointed out that, for Hamiltonian operators whose eigenvalues are known, which is often the case, this allows \textit{exact} solution of all Schr{\"o}dinger, von Neumann, and Heisenberg equations for finite quantum systems with unitary time evolution (the Heisenberg time evolution is also governed by $U(t,t_{0})$, but the solution for the expectation values of time-evolving observables is focused upon in that case, so $U(t,t_{0})$ is involved indirectly, yet still responsible for all dynamics).

For open quantum systems undergoing \textit{nonunitary} time evolution, these results can still be used to find explicit formulas for the $U(t,t_{0})$ of finite-sized effective joint systems, which can then be partial-traced over to get the nonunitary dynamics of the system of interest, provided that such a system is finite-dimensional.  While these results have already been worked out by the author, they are beyond the scope of the present discussion because they require much more space to explain.  Therefore, that will be the subject of future work.

We also showed several other useful results such as an exact finite-term BCH formula in \Sec{II.B}, and several examples of general operator decompositions in \Sec{II.C}.  Furthermore, we found a simple alternative form for elementary symmetric polynomials in \Eq{12}.

The main reason it is important to achieve symbolic forms, in any field of study, is that optimization typically requires derivatives, and having a symbolic formula permits differentiation, whereas that information becomes much harder to glean from strictly numerical results.

In closing, it is hoped that although the results of this paper are somewhat elementary, they can nevertheless prove useful in many applications, saving time and effort in otherwise tedious calculations for operator functions.  This may make tasks such as optimization of quantum systems significantly more approachable, and may even offer insights into many physical phenomena.
\begin{appendix}
\section{\label{sec:App.A}Derivation of Compact Matrix Inverse Formula}
From the Cayley-Hamilton theorem \cite[]{Ham1,Ham2,Ham3,Ham4,Cay1,Frob,Cay2}, given $n\times n$ matrix $A$ with characteristic equation
\begin{equation}
p(\lambda ) \equiv \det (\lambda I - A) \equiv \sum\limits_{k = 0}^n {c_k \lambda ^k }  = 0,
\label{eq:A.1}
\end{equation}
(unrelated to $p(n)$ of \Eq{17}), where expanding leads to a Vi{\`e}te's formula \cite[]{Viet,Gira,Newt},
\begin{equation}
c_k  = ( - 1)^{n - k} e_{n - k} (\bm{\lambda}' ),\;\;\;k \in 0, \ldots ,n,
\label{eq:A.2}
\end{equation}
where $e_{j} (\bm{\lambda} ')$ are elementary symmetric polynomials as in \Eq{11} and $\bm{\lambda}'\equiv(\lambda_{1}',\ldots,\lambda_{n}')$ are all eigenvalues of $A$ including repetitions, the Cayley-Hamilton theorem's main result is that \Eq{A.1} also holds with matrix argument $A$ as
\begin{equation}
p(A) = \sum\limits_{k = 0}^n {c_k A^k }  = 0I.
\label{eq:A.3}
\end{equation}
Then, in the rightmost equation of \Eq{A.3}, pulling out the first term, using $c_0  = ( - 1)^n \det (A)$ from \Eq{A.2}, multiplying through by $A^{-1}$, and solving for $A^{-1}$ gives
\begin{equation}
A^{ - 1}  = \frac{\text{adj}(A)}{\det(A)};\;\;\;\text{adj}(A)\equiv ( - 1)^{n + 1} \sum\limits_{k = 1}^n {c_k A^{k - 1} } ,
\label{eq:A.4}
\end{equation}
which defines the adjugate of $A$ as $\text{adj}(A)$, as seen in \Eq{5}.

While Vi{\`e}te's formula would give the $c_k$ in terms of $\bm{\lambda}'$, we can get $c_k$ directly from $A$ using discrete Fourier orthonormality.  Recall that the $n$-level discrete Fourier unitary matrix \cite[]{NiCh} has elements
\begin{equation}
F_{j,l}  \equiv F_{j,l}^{[n]}  = {\textstyle{1 \over {\sqrt n }}}e^{ - i{\textstyle{{2\pi } \over n}}(j - 1)(l - 1)}\;\; \text{for}\;\;j,l \in 1, \ldots ,n.
\label{eq:A.5}
\end{equation}
Like all unitaries, since the Hermitian conjugate is the inverse $F^{\dag}=F^{-1}$, we get orthonormality,
\begin{equation}
\sum\limits_{j = 1}^n {F_{j,k} ^* F_{j,l} }  = \delta _{k,l} .
\label{eq:A.6}
\end{equation}
To relate the $c_k$ to $A$ itself, we will use \Eq{A.1}, which has $n+1$ terms, so put $n\to n+1$ in \Eq{A.5} and \Eq{A.6}, and shift the index as $j\to j+1$, which yields
\begin{equation}
{\textstyle{1 \over {n + 1}}}\sum\limits_{j = 0}^n {e^{i{\textstyle{{2\pi } \over {n + 1}}}jk} e^{ - i{\textstyle{{2\pi } \over {n + 1}}}jl} }  = \delta _{k,l} .
\label{eq:A.7}
\end{equation}
Now, if we put $\lambda  \!=\! e^{ - i2\pi j/(n+1)} $ in \Eq{A.1} as $p(e^{ - i2\pi j/(n+1)})$ with new summation index $h$, and then multiply through by $e^{i2\pi jk/(n+1)}$, summing over $j$ from $0$ to $n$, and scaling by $\frac{1}{n+1}$ lets us use \Eq{A.7} to get
\begin{equation}
\begin{array}{*{20}r}
   {\sum\limits_{h = 0}^n {c_h e^{ - i{\textstyle{{2\pi } \over {n + 1}}}jh} } =} &\!\! {\det (e^{ - i{\textstyle{{2\pi } \over {n + 1}}}j} I - A)}  \\
   {e^{i{\textstyle{{2\pi } \over {n + 1}}}jk} \sum\limits_{h = 0}^n {c_h e^{ - i{\textstyle{{2\pi } \over {n + 1}}}jh} } =} &\!\! {e^{i{\textstyle{{2\pi } \over {n + 1}}}jk} p(e^{ - i{\textstyle{{2\pi } \over {n + 1}}}j} )}  \\
   {\sum\limits_{h = 0}^n {c_h \delta _{k,h} } =} &\!\! {\frac{1}{{n + 1}}\sum\limits_{j = 0}^n {e^{i{\textstyle{{2\pi } \over {n + 1}}}jk} p(e^{ - i{\textstyle{{2\pi } \over {n + 1}}}j} )} }  \\
   {c_k =} &\!\! {\frac{1}{{n + 1}}\sum\limits_{j = 0}^n {e^{i{\textstyle{{2\pi } \over {n + 1}}}jk} p(e^{ - i{\textstyle{{2\pi } \over {n + 1}}}j} )} ,}  \\
\end{array}
\label{eq:A.8}
\end{equation}
where $p(e^{ - i{\textstyle{{2\pi } \over {n + 1}}}j} ) = \det (e^{ - i{\textstyle{{2\pi } \over {n + 1}}}j} I - A)$, which gives the result in \Eq{6}.  Although there are many other ways to get the $c_k$, this way is particularly compact and useful.

Thus, \Eq{A.8} allows direct computation of $\text{adj}(A)$ in \Eq{A.4}, and together with the well-known formulas for the determinant and general Levi-Civita symbol given in \Eq{7} and \Eq{8}, \Eq{A.4} then yields the final closed-form result for $A^{-1}$ in \Eq{4} (written in terms of matrix $V$ there), where $A^{-1}$ exists if and only if (iff) $\det(A)\neq 0$.

Thus, \Eq{4} can be used for any nonsingular $n\times n$ matrix $V$, not just confluent Vandermondes.  While this result may seem trivially simple since algorithms are already well-known for the matrix inverse, keep in mind that \textit{many} published papers have labored over the inverse of the confluent Vandermonde \cite[]{Gaut,HoPa,LuRo,HoHo,SoMo}, with the simplest results still taking many pages to explain, whereas here we have achieved it in a single equation simply by deriving a more compact formula for an arbitrary matrix inverse.
\section{\label{sec:App.B}Inverse Confluent Vandermonde in Terms of Eigenvalues of its Parent Matrix}
First, to get the inverse confluent Vandermonde matrix $V^{-1}$ in terms of the \textit{elements} of $V$ only (and not in terms of $V$ as a whole, as in \Eq{4}), first we need
\begin{equation}
\begin{array}{*{20}l}
   {V^{k - 1}=} &\!\! {\sum\limits_{l,m = 1,1}^{n,n} \!\!{\left({ \,\sum\limits_{\mathbf{r} = \mathbf{1}}^{\mathbf{n}} {\delta_{r_{1},l}\delta_{r_{k},m}\prod\limits_{u = 1}^{k - 1} {V_{r_u ,r_{u + 1} } } }\!}\right)\! |l\rangle \langle m|,}}  \\
\end{array}
\label{eq:B.1}
\end{equation}
with vector indices $\mathbf{r} \equiv (r_1 , \ldots ,r_{k} )$, $\mathbf{1} \equiv (1_1 , \ldots ,1_{k} )$, $\mathbf{n} \equiv (n_1 , \ldots ,n_{k} )$, where $1_j  \equiv 1$, $n_j  \equiv n$, and $\{ |1\rangle , \ldots ,|n\rangle \}$ forms the complete orthonormal standard basis, and $\delta _{a,b}$ is the Kronecker delta.  Next, using
\begin{equation}
(e^{ - i{\textstyle{{2\pi } \over {n + 1}}}j} I - V)_{z,h_z }  = (e^{ - i{\textstyle{{2\pi } \over {n + 1}}}j} \delta _{z,h_z }  - V_{z,h_z } ),
\label{eq:B.2}
\end{equation}
where $z\in 1,\ldots n$ and $h_{z}\in 1,\ldots n$, together with \Eq{B.1} and \Eq{7} in \Eq{4} gives $V^{-1}$ in explicitly terms of $V_{a,b}$.

To get $V^{-1}$ explicitly in terms of the $M$ distinct eigenvalues $\bm{\lambda}\equiv(\lambda_{1},\ldots,\lambda_{M})$ of its parent matrix $A$ (the matrix whose eigenvalues define $V$ as given in \Sec{I}), we can either use \Eq{3} in \Eq{B.1} and \Eq{B.2}, or we can adapt \Eq{3} as
\begin{equation}
V_{c_{\alpha,\beta},d}={\kern -1pt}\frac{{(d - 1)!}\lambda _{\alpha} ^{d - \beta }\text{sgn}(\delta_{d,\beta}\!+\!\text{max}{\{0,d-\beta\}})}{[(d - \beta)\text{sgn}(\delta_{d,\beta}\!+\!\text{max}{\{0,d-\beta\}})]!},
\label{eq:B.3}
\end{equation}
where $\alpha \in 1, \ldots ,M$, $\beta \in 1, \ldots ,\mu _{\alpha}$, $d \in 1, \ldots ,n$, and $c_{\alpha ,\beta }$ is defined after \Eq{2}.  

However, in \Eqs{B.1}{B.3} and \Eq{4}, explicit calculation of elements of $V$ requires the \textit{inverse indical register function} for $c_{\alpha ,\beta }$, meaning: given some value $v$ in the range of $c_{\alpha ,\beta }$, what are $\alpha$ and $\beta$?  Thus, given $v\in 1,\ldots,n$, the values of $\alpha$ and $\beta$ for which \smash{$c_{\alpha ,\beta }  \equiv (\sum\nolimits_{j = 1}^{\alpha  - 1} {\mu _j })+\beta =v$} for $\alpha  \in 1, \ldots ,M$ and $\beta  \in 1, \ldots ,\mu _\alpha$ are
\begin{equation}
\alpha  \!=\!R_{v}\!\equiv\!\! \sum\limits_{a,b = 1,1}^{M,\max \{ \bm{\mu} \} }\!\!\!\!\!\!\! {a\delta _{v,c_{a,b} } } \;\;\;\text{and}\;\;\;\beta  \!=\!C_{v}\!\equiv\!\! \sum\limits_{a,b = 1,1}^{M,\max \{ \bm{\mu} \} }\!\!\!\!\!\!\! {b\delta _{v,c_{a,b} } },
\label{eq:B.4}
\end{equation}
where $c_{a ,b }  \equiv (\,\sum\nolimits_{j = 1}^{a  - 1} {\mu _j })+b$ and $\bm{\mu}\equiv(\mu_{1},\ldots,\mu_{M})$ as in \Sec{I}.  Thus, \Eq{B.4} transforms \Eq{B.3} to
\begin{equation}
V_{q,d}={\kern -1pt}\frac{{(d - 1)!}S_{d,C_{q}}}{{[(d - C_{q})S_{d,C_{q}}]!}}\lambda _{R_{q}} ^{d - C_{q} },
\label{eq:B.5}
\end{equation}
for $q,d\in 1,\ldots,n$, with abbreviating function,
\begin{equation}
S_{z,y}  \equiv {\mathop{\rm sgn}} (\delta _{z,y}  + \max \{ 0,z - y\} ) = \left\{ {\begin{array}{*{20}l}
   0; &\; {z < y}  \\
   1; &\; {z \ge y,}  \\
\end{array}} \right.
\label{eq:B.6}
\end{equation}
so now \Eq{B.5} gives elements of $V$ as a single-term function with no restrictions on its indices.

Putting everything together, we can now write $V^{-1}$ explicitly in terms of the distinct eigenvalues of its parent matrix $A$ by putting \Eq{B.5} into \Eq{B.1}, \Eq{B.2}, and \Eq{7}, and putting all of those into \Eq{4}, to get
\begin{widetext}
\vspace{-5pt}
\begin{equation}
V^{ - 1}  = \frac{{\left(\! \begin{array}{l}
 \frac{{( - 1)^{n + 1} }}{{n + 1}}\sum\limits_{k = 1}^n {\left[ {\sum\limits_{l,m = 1,1}^{n,n} {\left({\sum\limits_{\mathbf{r} = \mathbf{1}}^{\mathbf{n}} {\delta_{r_{1},l}\delta_{r_{k},m}\prod\limits_{u = 1}^{k - 1} {\frac{{(r_{u + 1}  - 1)! S_{r_{u + 1},C_{r_u} }\lambda _{R_{r_u } }^{r_{u + 1}  - C_{r_u} } }}{{[(r_{u + 1}  - C_{r_u})S_{r_{u+1},C_{r_u}} ]!}}} } }\right)|l\rangle \langle m|} } \right]}  \\ 
  \times \sum\limits_{j = 0}^n {e^{i\frac{{2\pi }}{{n + 1}}jk} } \sum\limits_{\mathbf{h} = \mathbf{1}}^{\mathbf{n}} {\left[ {\left( {\prod\limits_{x = 1,y = x + 1}^{n - 1,n}\!\!\!\!\!\!\!\!\! {{\mathop{\rm sgn}} (h_y  - h_x )} } \right)\!\left( {\prod\limits_{z = 1}^n {\!(e^{ - i\frac{{2\pi }}{{n + 1}}j} \delta _{z,h_z }  - \frac{{(h_{z}  - 1)! S_{h_{z},C_{z} }\lambda _{R_{z } }^{h_{z}  - C_{z} } }}{{[(h_{z}  - C_{z})S_{h_{z},C_{z}} ]!}})} \!} \right)} \right]}  \\ 
 \end{array}\! \right)_{\,_{\,^{\,_{\,_{\,_{\,}}}}}\!\!\!\!\!\!\!}}}{{\sum\limits_{\mathbf{k} = \mathbf{1}}^{\mathbf{n}} {\left[ {\left( {\prod\limits_{a = 1,b = a + 1}^{n - 1,n}\!\!\!\!\!\!\!\!\! {{\mathop{\rm sgn}} (k_b  - k_a )} } \right)\!\left( {\prod\limits_{c = 1}^n {\frac{{(k_{c}  - 1)! S_{k_{c},C_{c} }\lambda _{R_{c } }^{k_{c}  - C_{c} } }}{{[(k_{c} - C_{c})S_{k_{c},C_{c}} ]!}}} } \right)} \right]\rule{0pt}{19.0pt}} }},
\label{eq:B.7}
\end{equation}
\vspace{-5pt}
\end{widetext}
where $i\equiv\sqrt{-1}$, $\mathbf{r} \equiv (r_1 , \ldots ,r_{k} )$, and $\{ |1\rangle , \ldots ,|n\rangle \}$ forms the complete orthonormal standard basis as in \Eq{B.1}, and $\mathbf{h} \equiv (h_1 , \ldots ,h_n )$, $\mathbf{k} \equiv (k_1 , \ldots ,k_n )$, and $R_v$ and $C_v$ are given by \Eq{B.4}, $S_{z,y}$ is given by \Eq{B.6}, and $\mathbf{1}\equiv(1,\ldots,1)$ and $\mathbf{n}\equiv(n,\ldots,n)$ have dimensions $k$ or $n$\hsp{-0.2} depending\hsp{-0.2} on\hsp{-0.2} their\hsp{-0.2} usage.  Thus, \Eq{B.7} gives the inverse confluent Vandermonde $V^{-1}$ as an explicit function of the distinct eigenvalues of parent matrix $A$, in a single, nonrecursive equation that works for all multiplicity cases.
\section{\label{sec:App.C}Symbolic-Software-Friendly Inverse}
Adapting the results from \cite[]{Macd}, given $\mathbf{x} = (x_1 , \ldots ,x_m )$, and $m\times m$ matrix $X$ with full set of eigenvalues $\mathbf{x}$ including repetitions, if we define the $j\times j$ matrix
\begin{equation}
Q^{[j]}  \!\equiv\! \delta _{j,0}  \!+ \!\!\left( {\sum\limits_{a = 2}^j {\!(a \!-\! 1)|a \!-\! 1\rangle \langle a|} }\! \right)\!\! + \!\sum\limits_{q = 1}^j {p_q \!\sum\limits_{b = q}^j {\!|b\rangle \langle b\! -\! q \!+\! 1|,} }
\label{eq:C.1}
\end{equation}
where $\{ |1\rangle , \ldots ,|j\rangle \}$ is the standard basis and $p_q  \equiv p_q (\mathbf{x} ) \equiv \sum\nolimits_{c = 1}^m {x_c^{\shiftmath{0.5pt}{q}} } $ are power sums, then two useful forms of the elementary symmetric polynomials are
\begin{equation}
\begin{array}{*{20}c}
   {e_j (\mathbf{x})\equiv {\textstyle{1 \over {j!}}}\det (Q^{[j]} (\mathbf{x}))} & {\;\text{or}\;} & {e_j (X)\equiv {\textstyle{1 \over {j!}}}\det (Q^{[j]} (X)),}  \\
\end{array}
\label{eq:C.2}
\end{equation}
with $Q^{[j]} (\mathbf{x}) \!\equiv\! Q^{[j]} (\{p_q (\mathbf{x})\})$ and $Q^{[j]} (X) \!\equiv\! Q^{[j]} (\{p_q (X)\})$ and depending on the argument, we write the $p_q$ as
\begin{equation}
\begin{array}{*{20}c}
   {p_q (\mathbf{x})\equiv \sum\nolimits_{c = 1}^m {x_c^{\shiftmath{0.5pt}{q}} }} & {\;\text{or}\;} & {p_q (X)\equiv \text{tr}(X^q ).\rule{0pt}{9.5pt}}  \\
\end{array}
\label{eq:C.3}
\end{equation}
Thus, $e_j (\mathbf{x})$ in \Eq{C.2} provides a symbolic-software-friendly definition for the elementary symmetric polynomials since it has no complex-valued coefficients or constrained indices.  However, the structure of $Q^{[j]}$ will require many factors of $S_{z,y}$ from \Eq{B.6}, so \Eq{C.2} may not be as easy as \Eq{12} to use in theoretical calculations.

Then, using the matrix-argument form of \Eq{C.2} in \Eq{A.2} and \Eq{A.4} adapted to the present variables, we obtain
\begin{equation}
X^{ - 1}=\frac{1}{\det (X)}\sum\limits_{k = 1}^m {\frac{ \det [Q^{[m - k]} (X)]}{(m - k)!}(-X)^{k - 1} } ,
\label{eq:C.4}
\end{equation}
which is the symbolic-software-friendly alternative to \Eq{4} for the inverse of $m\times m$ matrix $X$, iff $\det(X)\neq 0$.

While \Eq{C.4} contains no complex-exponential factors (which is what makes it better for use with symbolic software), or index constraints, \Eq{4} is still preferable for certain theoretical calculations such as finding the inverse of the confluent Vandermonde matrix in terms of the eigenvalues of its parent matrix $A$, because again the structure of $Q^{[j]}$ introduces many more factors of $S_{z,y}$, while \Eq{4} becomes no more complicated than \Eq{B.7}, making \Eq{4} superior to \Eq{C.4} for certain applications.

Note that both \Eq{4} and \Eq{C.4} compute inverses explicitly in terms of elements of the input matrix or powers of its trace of powers, \textit{but they do this without any constraints on the indices}, which is something that has not been done until now, since the trace-based result of the Cayley-Hamilton theorem involves Bell polynomials which have complicated constraints on the indices.
\section{\label{sec:App.D}Derivation of Compact Elementary Symmetric Polynomials Formula}
To get a better formula for the elementary symmetric polynomials than the traditional form in \Eq{11}, first invert \Eq{A.2} and adapt for general input $\mathbf{x} = (x_1 , \ldots ,x_m )$ as
\begin{equation}
e_j (\mathbf{x}) = ( - 1)^j c_{m - j} ,\;\;\;j \in 0, \ldots ,m.
\label{eq:D.1}
\end{equation}
From \Eq{A.8}, the characteristic polynomial coefficients for an $m$-level matrix $X$ with full set of eigenvalues $\mathbf{x}$ including repetitions, with index adapted for \Eq{D.1}, are
\begin{equation}
c_{m - j}  = \frac{1}{{m + 1}}\sum\limits_{k = 0}^m {e^{i\frac{{2\pi }}{{m + 1}}k(m - j)} \det (e^{ - i\frac{{2\pi }}{{m + 1}}k} I - X)}.
\label{eq:D.2}
\end{equation}
Since this holds for any $X$ with eigenvalues $\mathbf{x}$, we can choose the simplest one as $X\equiv\text{diag}(\mathbf{x})$, which gives
\begin{equation}
\det (e^{ - i\frac{{2\pi }}{{m + 1}}k} I - X) = \prod\limits_{l = 1}^m {(e^{ - i\frac{{2\pi }}{{m + 1}}k}  - x_l )}.
\label{eq:D.3}
\end{equation}
Putting \Eq{D.3} into \Eq{D.2} and putting that into \Eq{D.1} gives
\begin{equation}
e_j (\mathbf{x}) = \frac{{( - 1)^j }}{{m + 1}}\sum\limits_{k = 0}^m {e^{i\frac{{2\pi }}{{m + 1}}k(m - j)} \prod\limits_{l = 1}^m {(e^{ - i\frac{{2\pi }}{{m + 1}}k}  - x_l )} } ,
\label{eq:D.4}
\end{equation}
for $j \in 0, \ldots ,m$ where $m\equiv\text{dim}(\mathbf{x})$, which is the result in \Eq{12}, thus avoiding the need for nested sums with related constrained indices as in \Eq{11}.
\end{appendix}
%

\begin{thebibliography}{38}%
\makeatletter
\providecommand \@ifxundefined [1]{%
 \@ifx{#1\undefined}
}%
\providecommand \@ifnum [1]{%
 \ifnum #1\expandafter \@firstoftwo
 \else \expandafter \@secondoftwo
 \fi
}%
\providecommand \@ifx [1]{%
 \ifx #1\expandafter \@firstoftwo
 \else \expandafter \@secondoftwo
 \fi
}%
\providecommand \natexlab [1]{#1}%
\providecommand \enquote  [1]{``#1''}%
\providecommand \bibnamefont  [1]{#1}%
\providecommand \bibfnamefont [1]{#1}%
\providecommand \citenamefont [1]{#1}%
\providecommand \href@noop [0]{\@secondoftwo}%
\providecommand \href [0]{\begingroup \@sanitize@url \@href}%
\providecommand \@href[1]{\@@startlink{#1}\@@href}%
\providecommand \@@href[1]{\endgroup#1\@@endlink}%
\providecommand \@sanitize@url [0]{\catcode `\\12\catcode `\$12\catcode
  `\&12\catcode `\#12\catcode `\^12\catcode `\_12\catcode `\%12\relax}%
\providecommand \@@startlink[1]{}%
\providecommand \@@endlink[0]{}%
\providecommand \url  [0]{\begingroup\@sanitize@url \@url }%
\providecommand \@url [1]{\endgroup\@href {#1}{\urlprefix }}%
\providecommand \urlprefix  [0]{URL }%
\providecommand \Eprint [0]{\href }%
\providecommand \doibase [0]{http://dx.doi.org/}%
\providecommand \selectlanguage [0]{\@gobble}%
\providecommand \bibinfo  [0]{\@secondoftwo}%
\providecommand \bibfield  [0]{\@secondoftwo}%
\providecommand \translation [1]{[#1]}%
\providecommand \BibitemOpen [0]{}%
\providecommand \bibitemStop [0]{}%
\providecommand \bibitemNoStop [0]{.\EOS\space}%
\providecommand \EOS [0]{\spacefactor3000\relax}%
\providecommand \BibitemShut  [1]{\csname bibitem#1\endcsname}%
\let\auto@bib@innerbib\@empty
\bibitem [{Ham(1853)}]{Ham1}%
  \BibitemOpen
  \href@noop {} {\emph {\bibinfo {title} {Lectures on Quaternions}}}\ (\bibinfo
  {address} {Royal Irish Academy, Dublin},\ \bibinfo {year} {1853})\BibitemShut
  {NoStop}%
\bibitem [{\citenamefont {Hamilton}(1862{\natexlab{a}})}]{Ham2}%
  \BibitemOpen
  \bibfield  {author} {\bibinfo {author} {\bibfnamefont {W.~R.}\ \bibnamefont
  {Hamilton}},\ }in\ \href@noop {} {\emph {\bibinfo {booktitle} {Proc. Roy.
  Irish. Acad.}}},\ Vol.\ \bibinfo {volume} {viii}\ (\bibinfo {year} {1862})\
  p.\ \bibinfo {pages} {182}\BibitemShut {NoStop}%
\bibitem [{\citenamefont {Hamilton}(1862{\natexlab{b}})}]{Ham3}%
  \BibitemOpen
  \bibfield  {author} {\bibinfo {author} {\bibfnamefont {W.~R.}\ \bibnamefont
  {Hamilton}},\ }in\ \href@noop {} {\emph {\bibinfo {booktitle} {Proc. Roy.
  Irish. Acad.}}},\ Vol.\ \bibinfo {volume} {viii}\ (\bibinfo {year} {1862})\
  p.\ \bibinfo {pages} {190}\BibitemShut {NoStop}%
\bibitem [{\citenamefont {Hamilton}(1862{\natexlab{c}})}]{Ham4}%
  \BibitemOpen
  \bibfield  {author} {\bibinfo {author} {\bibfnamefont {W.~R.}\ \bibnamefont
  {Hamilton}},\ }\href@noop {} {\bibfield  {journal} {\bibinfo  {journal} {The
  London, Edinburgh and Dublin Philosophical Magazine and Journal of Science}\
  }\textbf {\bibinfo {volume} {iv}} (\bibinfo {year}
  {1862}{\natexlab{c}})}\BibitemShut {NoStop}%
\bibitem [{\citenamefont {Cayley}(1858)}]{Cay1}%
  \BibitemOpen
  \bibfield  {author} {\bibinfo {author} {\bibfnamefont {A.}~\bibnamefont
  {Cayley}},\ }\href@noop {} {\bibfield  {journal} {\bibinfo  {journal}
  {Philos. Trans. R. Soc. London}\ }\textbf {\bibinfo {volume} {148}},\
  \bibinfo {pages} {17} (\bibinfo {year} {1858})}\BibitemShut {NoStop}%
\bibitem [{\citenamefont {Frobenius}(1878)}]{Frob}%
  \BibitemOpen
  \bibfield  {author} {\bibinfo {author} {\bibfnamefont {G.}~\bibnamefont
  {Frobenius}},\ }\href@noop {} {\bibfield  {journal} {\bibinfo  {journal}
  {Journal f{\"u}r die reine und angewandte Mathematik}\ }\textbf {\bibinfo
  {volume} {84}},\ \bibinfo {pages} {1} (\bibinfo {year} {1878})}\BibitemShut
  {NoStop}%
\bibitem [{\citenamefont {Cayley}(1889)}]{Cay2}%
  \BibitemOpen
  \bibfield  {author} {\bibinfo {author} {\bibfnamefont {A.}~\bibnamefont
  {Cayley}},\ }\href@noop {} {\emph {\bibinfo {title} {The Collected
  Mathematical Papers of Arthur Cayley}}}\ (\bibinfo  {publisher} {Cambridge
  University Press},\ \bibinfo {year} {1889})\ p.\ \bibinfo {pages}
  {475}\BibitemShut {NoStop}%
\bibitem [{\citenamefont {Gautschi}(1962)}]{Gaut}%
  \BibitemOpen
  \bibfield  {author} {\bibinfo {author} {\bibfnamefont {W.}~\bibnamefont
  {Gautschi}},\ }\href@noop {} {\bibfield  {journal} {\bibinfo  {journal}
  {Numerische Mathematik}\ }\textbf {\bibinfo {volume} {4}},\ \bibinfo {pages}
  {117} (\bibinfo {year} {1962})}\BibitemShut {NoStop}%
\bibitem [{\citenamefont {Hou}\ and\ \citenamefont {Pang}(2002)}]{HoPa}%
  \BibitemOpen
  \bibfield  {author} {\bibinfo {author} {\bibfnamefont {S.~H.}\ \bibnamefont
  {Hou}}\ and\ \bibinfo {author} {\bibfnamefont {W.~K.}\ \bibnamefont {Pang}},\
  }\href@noop {} {\bibfield  {journal} {\bibinfo  {journal} {Comp. Math. App.}\
  }\textbf {\bibinfo {volume} {43}},\ \bibinfo {pages} {1539} (\bibinfo {year}
  {2002})}\BibitemShut {NoStop}%
\bibitem [{\citenamefont {Luther}\ and\ \citenamefont {Rost}(2004)}]{LuRo}%
  \BibitemOpen
  \bibfield  {author} {\bibinfo {author} {\bibfnamefont {U.}~\bibnamefont
  {Luther}}\ and\ \bibinfo {author} {\bibfnamefont {K.}~\bibnamefont {Rost}},\
  }\href@noop {} {\bibfield  {journal} {\bibinfo  {journal} {Elec. Trans. Num.
  An.}\ }\textbf {\bibinfo {volume} {18}},\ \bibinfo {pages} {91} (\bibinfo
  {year} {2004})}\BibitemShut {NoStop}%
\bibitem [{\citenamefont {Hou}\ and\ \citenamefont {Hou}(2007)}]{HoHo}%
  \BibitemOpen
  \bibfield  {author} {\bibinfo {author} {\bibfnamefont {S.~H.}\ \bibnamefont
  {Hou}}\ and\ \bibinfo {author} {\bibfnamefont {E.}~\bibnamefont {Hou}},\
  }\href@noop {} {\bibfield  {journal} {\bibinfo  {journal} {Elec. J. Math.
  Tech.}\ }\textbf {\bibinfo {volume} {1}},\ \bibinfo {pages} {11} (\bibinfo
  {year} {2007})}\BibitemShut {NoStop}%
\bibitem [{\citenamefont {Soto-Eguibar}\ and\ \citenamefont
  {Moya-Cessa}(2011)}]{SoMo}%
  \BibitemOpen
  \bibfield  {author} {\bibinfo {author} {\bibfnamefont {F.}~\bibnamefont
  {Soto-Eguibar}}\ and\ \bibinfo {author} {\bibfnamefont {H.}~\bibnamefont
  {Moya-Cessa}},\ }\href@noop {} {\bibfield  {journal} {\bibinfo  {journal}
  {App. Math. Inf. Sci.}\ }\textbf {\bibinfo {volume} {5}},\ \bibinfo {pages}
  {361} (\bibinfo {year} {2011})}\BibitemShut {NoStop}%
\bibitem [{\citenamefont {Schr{\"o}dinger}(1926{\natexlab{a}})}]{Sch1}%
  \BibitemOpen
  \bibfield  {author} {\bibinfo {author} {\bibfnamefont {E.}~\bibnamefont
  {Schr{\"o}dinger}},\ }\href@noop {} {\bibfield  {journal} {\bibinfo
  {journal} {Annelen der Physik}\ }\textbf {\bibinfo {volume} {384}},\ \bibinfo
  {pages} {361} (\bibinfo {year} {1926}{\natexlab{a}})}\BibitemShut {NoStop}%
\bibitem [{\citenamefont {Schr{\"o}dinger}(1926{\natexlab{b}})}]{Sch2}%
  \BibitemOpen
  \bibfield  {author} {\bibinfo {author} {\bibfnamefont {E.}~\bibnamefont
  {Schr{\"o}dinger}},\ }\href@noop {} {\bibfield  {journal} {\bibinfo
  {journal} {Phys. Rev.}\ }\textbf {\bibinfo {volume} {28}},\ \bibinfo {pages}
  {1049} (\bibinfo {year} {1926}{\natexlab{b}})}\BibitemShut {NoStop}%
\bibitem [{\citenamefont {Ricci}\ and\ \citenamefont
  {Levi-Civita}(1901)}]{RiLe}%
  \BibitemOpen
  \bibfield  {author} {\bibinfo {author} {\bibfnamefont {M.~M.~G.}\
  \bibnamefont {Ricci}}\ and\ \bibinfo {author} {\bibfnamefont
  {T.}~\bibnamefont {Levi-Civita}},\ }\href@noop {} {\bibfield  {journal}
  {\bibinfo  {journal} {Mathematische Annalen}\ }\textbf {\bibinfo {volume}
  {54}},\ \bibinfo {pages} {125} (\bibinfo {year} {1901})}\BibitemShut
  {NoStop}%
\bibitem [{\citenamefont {Moler}\ and\ \citenamefont {Loan}(1978)}]{MoV1}%
  \BibitemOpen
  \bibfield  {author} {\bibinfo {author} {\bibfnamefont {C.}~\bibnamefont
  {Moler}}\ and\ \bibinfo {author} {\bibfnamefont {C.~V.}\ \bibnamefont
  {Loan}},\ }\href@noop {} {\bibfield  {journal} {\bibinfo  {journal} {SIAM
  Rev.}\ }\textbf {\bibinfo {volume} {20}},\ \bibinfo {pages} {801} (\bibinfo
  {year} {1978})}\BibitemShut {NoStop}%
\bibitem [{\citenamefont {Moler}\ and\ \citenamefont {Loan}(2003)}]{MoV2}%
  \BibitemOpen
  \bibfield  {author} {\bibinfo {author} {\bibfnamefont {C.}~\bibnamefont
  {Moler}}\ and\ \bibinfo {author} {\bibfnamefont {C.~V.}\ \bibnamefont
  {Loan}},\ }\href@noop {} {\bibfield  {journal} {\bibinfo  {journal} {SIAM
  Rev.}\ }\textbf {\bibinfo {volume} {45}},\ \bibinfo {pages} {3} (\bibinfo
  {year} {2003})}\BibitemShut {NoStop}%
\bibitem [{\citenamefont {Curtright}\ \emph {et~al.}(2014)\citenamefont
  {Curtright}, \citenamefont {Fairlie},\ and\ \citenamefont {Zachos}}]{CuFZ}%
  \BibitemOpen
  \bibfield  {author} {\bibinfo {author} {\bibfnamefont {T.~L.}\ \bibnamefont
  {Curtright}}, \bibinfo {author} {\bibfnamefont {D.~B.}\ \bibnamefont
  {Fairlie}}, \ and\ \bibinfo {author} {\bibfnamefont {C.~K.}\ \bibnamefont
  {Zachos}},\ }\href@noop {} {\bibfield  {journal} {\bibinfo  {journal}
  {SIGMA}\ }\textbf {\bibinfo {volume} {10}},\ \bibinfo {pages} {084} (\bibinfo
  {year} {2014})}\BibitemShut {NoStop}%
\bibitem [{\citenamefont {Curtright}(2015)}]{Curt}%
  \BibitemOpen
  \bibfield  {author} {\bibinfo {author} {\bibfnamefont {T.~L.}\ \bibnamefont
  {Curtright}},\ }\href@noop {} {\bibfield  {journal} {\bibinfo  {journal} {J.
  Math. Phys.}\ }\textbf {\bibinfo {volume} {56}},\ \bibinfo {pages} {091703}
  (\bibinfo {year} {2015})}\BibitemShut {NoStop}%
\bibitem [{\citenamefont {Euler}(1753)}]{Eule}%
  \BibitemOpen
  \bibfield  {author} {\bibinfo {author} {\bibfnamefont {L.}~\bibnamefont
  {Euler}},\ }\href@noop {} {\bibfield  {journal} {\bibinfo  {journal} {Novi
  Commentarii Academiae Scientiarum Petropolitanae}\ }\textbf {\bibinfo
  {volume} {3}},\ \bibinfo {pages} {125} (\bibinfo {year} {1753})}\BibitemShut
  {NoStop}%
\bibitem [{\citenamefont {Skiena}(1990)}]{Skie}%
  \BibitemOpen
  \bibfield  {author} {\bibinfo {author} {\bibfnamefont {S.}~\bibnamefont
  {Skiena}},\ }\href@noop {} {\emph {\bibinfo {title} {Implementing Discrete
  Mathematics: Combinatorics and Graph Theory with Mathematica}}}\ (\bibinfo
  {publisher} {Basic Books},\ \bibinfo {year} {1990})\ p.~\bibinfo {pages}
  {57}\BibitemShut {NoStop}%
\bibitem [{\citenamefont {Hellwig}\ and\ \citenamefont {Kraus}(1969)}]{HKr1}%
  \BibitemOpen
  \bibfield  {author} {\bibinfo {author} {\bibfnamefont {K.~E.}\ \bibnamefont
  {Hellwig}}\ and\ \bibinfo {author} {\bibfnamefont {K.}~\bibnamefont
  {Kraus}},\ }\href@noop {} {\bibfield  {journal} {\bibinfo  {journal} {Commun.
  Math. Phys.}\ }\textbf {\bibinfo {volume} {11}},\ \bibinfo {pages} {214}
  (\bibinfo {year} {1969})}\BibitemShut {NoStop}%
\bibitem [{\citenamefont {Hellwig}\ and\ \citenamefont {Kraus}(1970)}]{HKr2}%
  \BibitemOpen
  \bibfield  {author} {\bibinfo {author} {\bibfnamefont {K.~E.}\ \bibnamefont
  {Hellwig}}\ and\ \bibinfo {author} {\bibfnamefont {K.}~\bibnamefont
  {Kraus}},\ }\href@noop {} {\bibfield  {journal} {\bibinfo  {journal} {Commun.
  Math. Phys.}\ }\textbf {\bibinfo {volume} {16}},\ \bibinfo {pages} {142}
  (\bibinfo {year} {1970})}\BibitemShut {NoStop}%
\bibitem [{\citenamefont {Choi}(1975)}]{Choi}%
  \BibitemOpen
  \bibfield  {author} {\bibinfo {author} {\bibfnamefont {M.~D.}\ \bibnamefont
  {Choi}},\ }\href@noop {} {\bibfield  {journal} {\bibinfo  {journal} {Linear
  Algebra Appl.}\ }\textbf {\bibinfo {volume} {10}},\ \bibinfo {pages} {285}
  (\bibinfo {year} {1975})}\BibitemShut {NoStop}%
\bibitem [{\citenamefont {Kraus}(1983)}]{Kra1}%
  \BibitemOpen
  \bibfield  {author} {\bibinfo {author} {\bibfnamefont {K.}~\bibnamefont
  {Kraus}},\ }\href@noop {} {\emph {\bibinfo {title} {States, Effects, and
  Operations: Fundamental Notions of Quantum Theory}}}\ (\bibinfo  {publisher}
  {Springer-Verlag, Berlin},\ \bibinfo {year} {1983})\ p.\ \bibinfo {pages}
  {190}\BibitemShut {NoStop}%
\bibitem [{\citenamefont {Hedemann}(2014)}]{HedD}%
  \BibitemOpen
  \bibfield  {author} {\bibinfo {author} {\bibfnamefont {S.~R.}\ \bibnamefont
  {Hedemann}},\ }\emph {\bibinfo {title} {Hyperspherical Bloch Vectors with
  Applications to Entanglement and Quantum State Tomography}},\ \href@noop {}
  {Ph.D. thesis},\ \bibinfo  {school} {Stevens Institute of Technology}
  (\bibinfo {year} {2014}),\ \bibinfo {note} {~UMI Diss. Pub.
  3636036}\BibitemShut {NoStop}%
\bibitem [{\citenamefont {Sakurai}(1994)}]{Saku}%
  \BibitemOpen
  \bibfield  {author} {\bibinfo {author} {\bibfnamefont {J.~J.}\ \bibnamefont
  {Sakurai}},\ }\href@noop {} {\emph {\bibinfo {title} {Modern Quantum
  Mechanics, Revised Edition}}},\ edited by\ \bibinfo {editor} {\bibfnamefont
  {S.~F.}\ \bibnamefont {Tuan}}\ (\bibinfo  {publisher} {Addison-Wesley},\
  \bibinfo {year} {1994})\ pp.\ \bibinfo {pages} {72, 96}\BibitemShut {NoStop}%
\bibitem [{\citenamefont {Suzuki}(1993)}]{Suzu}%
  \BibitemOpen
  \bibfield  {author} {\bibinfo {author} {\bibfnamefont {M.}~\bibnamefont
  {Suzuki}},\ }\href@noop {} {\bibfield  {journal} {\bibinfo  {journal} {Proc.
  Japan Acad.}\ }\textbf {\bibinfo {volume} {69B}},\ \bibinfo {pages} {161}
  (\bibinfo {year} {1993})}\BibitemShut {NoStop}%
\bibitem [{\citenamefont {Gardiner}\ and\ \citenamefont
  {Collett}(1985)}]{GaCo}%
  \BibitemOpen
  \bibfield  {author} {\bibinfo {author} {\bibfnamefont {C.~W.}\ \bibnamefont
  {Gardiner}}\ and\ \bibinfo {author} {\bibfnamefont {M.~J.}\ \bibnamefont
  {Collett}},\ }\href@noop {} {\bibfield  {journal} {\bibinfo  {journal} {Phys.
  Rev. A}\ }\textbf {\bibinfo {volume} {31}},\ \bibinfo {pages} {3761}
  (\bibinfo {year} {1985})}\BibitemShut {NoStop}%
\bibitem [{\citenamefont {Gardiner}\ and\ \citenamefont {Zoller}(2010)}]{GaZo}%
  \BibitemOpen
  \bibfield  {author} {\bibinfo {author} {\bibfnamefont {C.~W.}\ \bibnamefont
  {Gardiner}}\ and\ \bibinfo {author} {\bibfnamefont {P.}~\bibnamefont
  {Zoller}},\ }\href@noop {} {\emph {\bibinfo {title} {Quantum Noise}}},\
  \bibinfo {edition} {3rd}\ ed.\ (\bibinfo  {publisher} {Springer-Verlag},\
  \bibinfo {year} {2010})\BibitemShut {NoStop}%
\bibitem [{\citenamefont {Giscard}\ \emph {et~al.}(2015)\citenamefont
  {Giscard}, \citenamefont {Lui}, \citenamefont {Thwaite},\ and\ \citenamefont
  {Jaksch}}]{GLTJ}%
  \BibitemOpen
  \bibfield  {author} {\bibinfo {author} {\bibfnamefont {P.~L.}\ \bibnamefont
  {Giscard}}, \bibinfo {author} {\bibfnamefont {K.}~\bibnamefont {Lui}},
  \bibinfo {author} {\bibfnamefont {S.~J.}\ \bibnamefont {Thwaite}}, \ and\
  \bibinfo {author} {\bibfnamefont {D.}~\bibnamefont {Jaksch}},\ }\href@noop {}
  {\bibfield  {journal} {\bibinfo  {journal} {J. Math. Phys.}\ }\textbf
  {\bibinfo {volume} {56}},\ \bibinfo {pages} {053503} (\bibinfo {year}
  {2015})}\BibitemShut {NoStop}%
\bibitem [{\citenamefont {Hedemann}(2013)}]{HedU}%
  \BibitemOpen
  \bibfield  {author} {\bibinfo {author} {\bibfnamefont {S.~R.}\ \bibnamefont
  {Hedemann}},\ }\href@noop {} {\bibfield  {journal} {\bibinfo  {journal}
  {arXiv quant-ph}\ }\textbf {\bibinfo {volume} {1303}},\ \bibinfo {pages}
  {5904} (\bibinfo {year} {2013})},\ \bibinfo {note}
  {\href{https://arxiv.org/abs/1303.5904}{arXiv:1303.5904}}\BibitemShut
  {NoStop}%
\bibitem [{\citenamefont {Bloch}(1946)}]{Blch}%
  \BibitemOpen
  \bibfield  {author} {\bibinfo {author} {\bibfnamefont {F.}~\bibnamefont
  {Bloch}},\ }\href@noop {} {\bibfield  {journal} {\bibinfo  {journal} {Phys.
  Rev.}\ }\textbf {\bibinfo {volume} {70}},\ \bibinfo {pages} {460} (\bibinfo
  {year} {1946})}\BibitemShut {NoStop}%
\bibitem [{\citenamefont {Nielsen}\ and\ \citenamefont {Chuang}(2010)}]{NiCh}%
  \BibitemOpen
  \bibfield  {author} {\bibinfo {author} {\bibfnamefont {M.~A.}\ \bibnamefont
  {Nielsen}}\ and\ \bibinfo {author} {\bibfnamefont {I.~L.}\ \bibnamefont
  {Chuang}},\ }\href@noop {} {\emph {\bibinfo {title} {Quantum Computation and
  Quantum Information}}}\ (\bibinfo  {publisher} {Cambridge University Press},\
  \bibinfo {year} {2010})\ pp.\ \bibinfo {pages} {15, 291, 216}\BibitemShut
  {NoStop}%
\bibitem [{\citenamefont {Vi{\`e}te}(1615)}]{Viet}%
  \BibitemOpen
  \bibfield  {author} {\bibinfo {author} {\bibfnamefont {F.}~\bibnamefont
  {Vi{\`e}te}},\ }\href@noop {} {\bibfield  {journal} {\bibinfo  {journal}
  {Opera Mathematica}\ }\textbf {\bibinfo {volume} {III}} (\bibinfo {year}
  {1615})}\BibitemShut {NoStop}%
\bibitem [{\citenamefont {Girard}(1629)}]{Gira}%
  \BibitemOpen
  \bibfield  {author} {\bibinfo {author} {\bibfnamefont {A.}~\bibnamefont
  {Girard}},\ }\href@noop {} {\emph {\bibinfo {title} {Invention Nouvelle en
  l'Alg{\`e}bre}}}\ (\bibinfo  {publisher} {Chez Guillaume Ianffon Blaeuw},\
  \bibinfo {year} {1629})\BibitemShut {NoStop}%
\bibitem [{\citenamefont {Newton}(1707)}]{Newt}%
  \BibitemOpen
  \bibfield  {author} {\bibinfo {author} {\bibfnamefont {I.}~\bibnamefont
  {Newton}},\ }\href@noop {} {\emph {\bibinfo {title} {Arithmetica
  Universalis}}}\ (\bibinfo  {publisher} {William Whiston},\ \bibinfo {year}
  {1707})\BibitemShut {NoStop}%
\bibitem [{\citenamefont {Macdonald}(1979)}]{Macd}%
  \BibitemOpen
  \bibfield  {author} {\bibinfo {author} {\bibfnamefont {I.~G.}\ \bibnamefont
  {Macdonald}},\ }\href@noop {} {\emph {\bibinfo {title} {Symmetric Functions
  and Hall Polynomials}}},\ \bibinfo {edition} {2nd}\ ed.\ (\bibinfo
  {publisher} {Oxford University Press},\ \bibinfo {year} {1979})\ p.~\bibinfo
  {pages} {28}\BibitemShut {NoStop}%
\end{thebibliography}
\end{document}